\documentclass[conference]{IEEEtran}
\IEEEoverridecommandlockouts
\usepackage{cite}
\usepackage{amsmath,amssymb,amsfonts}
\usepackage{algorithmic}
\usepackage{graphicx}
\usepackage{textcomp}
\usepackage{xcolor}
\def\BibTeX{{\rm B\kern-.05em{\sc i\kern-.025em b}\kern-.08em
    T\kern-.1667em\lower.7ex\hbox{E}\kern-.125emX}}

\usepackage{caption}
\usepackage{subcaption}

\usepackage{hyperref}
\hypersetup{
    colorlinks=true,
    linkcolor=blue,
    filecolor=magenta,      
    urlcolor=cyan,
}

\usepackage{multirow}    
\usepackage{tabularx}    
\usepackage{ragged2e}     
\usepackage{adjustbox}
\begin{document}

\title{Towards Fine-Grained Scalability for Stateful Stream Processing Systems}

\author{\IEEEauthorblockN{1\textsuperscript{st} Yunfan Qing}
\IEEEauthorblockA{\textit{School of Electronic Information \& Electrical Engineering} \\
\textit{Shanghai Jiao Tong University}\\
Shanghai, China \\
yfqing@sjtu.edu.cn}
\and
\IEEEauthorblockN{2\textsuperscript{nd} Wenli Zheng}
\IEEEauthorblockA{\textit{School of Electronic Information \& Electrical Engineering} \\
\textit{Shanghai Jiao Tong University}\\
Shanghai, China \\
zheng-wl@cs.sjtu.edu.cn}
}

\maketitle

\begin{abstract}
Dynamic scaling is critical to stream processing engines, as their long-running nature demands adaptive resource management.
Existing scaling approaches easily cause performance degradation due to coarse-grained synchronization and inefficient state migration, resulting in system halt or high processing latency.
In this paper, we propose DRRS, an on-the-fly scaling method that reduces
performance overhead at the system level with three key innovations: 
(i) fine-grained scaling signals coupled with a re-routing mechanism that significantly mitigates propagation delay, 
(ii) a sophisticated record-scheduling mechanism that substantially reduces processing suspension,
and (iii) subscale division, a mechanism that partitions migrating states into independent subsets, thereby reducing dependency-related overhead to enable finer-grained control and better runtime adaptability during scaling. 
DRRS is implemented on Apache Flink and, when compared to state-of-the-art approaches, reduces peak and average latencies by up to 81.1\% and 95.5\% respectively, while achieving a 72.8\%-86\% reduction in scaling duration, without disruption in non-scaling periods. 

\end{abstract}


\section{Introduction}
Stream processing engines (SPEs) \cite{carbone2015apache,murray2013naiad,zaharia2016apache,floratou2017dhalion,armbrust2018structured} are fundamental components in real-time data processing systems, supporting diverse applications from recommendation platforms to fraud detection and health monitoring\cite{carcillo2018scarff,noghabi2017samza,greco2019edge}.
The long-running nature and real-time processing requirements of SPEs necessitate dynamic scaling to adapt to fluctuating workloads.
While prior work has studied how to make \textbf{scaling decisions}\cite{zhang2021autrascale,floratou2017dhalion,buddhika2017online,de2017proactive,ma2018optimization}, which determine when and how to scale according to the system status and predefined policies, much fewer efforts have been made on improving \textbf{scaling mechanisms}. Scaling mechanisms determine how to execute scaling operations, focusing on systematically reducing performance degradation without changing processing results.
Scaling mechanisms in stateful SPEs requires delicate synchronization and coordination, particularly for state migration across multiple worker nodes\cite{carbone2017state,noghabi2017samza}, thus presenting significant challenges. 

Mainstream SPEs \cite{apache_flink_elastic_scaling,kulkarni2015twitter,armbrust2018structured} mainly employ the Stop-Checkpoint-Restart mechanism for dynamic scaling, 
which involves halting the system to create a checkpoint of the global state, followed by restarting with a new configuration based on this checkpoint. 
However, it can introduce substantial system downtime, rendering it unsuitable for latency-sensitive applications~\cite{shukla2018toward,gu2022meces}. 
Recent research has focused on on-the-fly dynamic scaling to avoid system halt \cite{gu2022meces,del2020rhino,hoffmann2019megaphone,shukla2018toward,mai2018chi,mao2021trisk,rajadurai2018gloss}, 
but three critical challenges still persist:
\begin{enumerate}
  \item \textbf{Propagation delays} of scaling signals, caused by {\color{black}long transmission paths and alignment requirements, create cumulative effects that compromise dynamic workload adaptation.} 
  \item \textbf{Processing suspensions} during state migration are necessary to ensure the state associated with an input's key is fully localized before processing, thereby significantly increasing processing latency.
  \item \textbf{Dependency-related overhead}, arising from the dependencies between state units during migration, has been underexplored in prior work. It not only introduces additional latency, but also constrains the control granularity and runtime adaptability.
\end{enumerate}

To address the challenges, we propose DRRS, an on-the-fly scaling method to mitigate performance degradation during scaling in stateful SPEs through three key innovations.
First, the \textbf{Decoupling and Re-routing} mechanism employs decoupled scaling signals with direct predecessor injection, enabling topologically shortest propagation paths and alignment-free state migration triggering.
Second, the \textbf{Record Scheduling} mechanism permits legitimate adjustments to the execution order---the specific sequence in which records are internally dispatched to operators---while ensuring semantic reservation, significantly reducing processing suspensions during state migration. 
Third, the \textbf{Subscale Division} mechanism partitions the scaling process into independent subscales, each managing a subset of state units 
without additional setup overhead.
It reduces dependency-related overhead and enables finer-grained control and better runtime adaptability.
DRRS ensures the scaling correctness with fully preserved execution semantics. Its output is identical to that of a non-scaling execution for deterministic operators, or one of the valid outcomes with acceptable guarantee for non-deterministic operators.

DRRS\footnote{The source code is available at \url{https://github.com/floudk/DRRS}} is implemented as a pluggable module, to be seamlessly integrated with existing SPEs without performance degradation in non-scaling periods. 
Evaluations using diverse benchmarks and real-world datasets show that {\color{black}it achieves up to 81.1\% and 95.5\% reductions in peak and average latencies, while accelerating scaling by 72.8\%-86\%} compared to state-of-the-art approaches.
To our knowledge, DRRS is the first method 
to introduce fine-grained synchronization and {\color{black} record-scheduling mechanisms in stateful SPE scaling,} and also the first to achieve dynamically schedulable scaling operations. To summarize, this paper makes the following major contributions: 
\begin{itemize}
\item We propose a fine-grained synchronization mechanism that 
mitigates propagation delays with decoupled scaling signals.

\item We propose 
a {\color{black}record-scheduling} mechanism that reduces processing suspensions by allowing legitimate adjustments to the {\color{black}engine's internal record-execution order} while preserving semantics during scaling.

\item We introduce a 
subscale division mechanism that 
reduces migration dependency overhead and enables parallelizable and schedulable scaling. 

\item We implement DRRS as a plugin on Apache Flink, requiring no user code modifications and causing no disruption during non-scaling periods. 
\end{itemize}

The remainder of this paper is organized as follows. 
Section \ref{background} presents background and motivation for this work. 
Sections \ref{design} and \ref{implementation} detail the design and implementation of DRRS.
Section \ref{evaluation} evaluates DRRS, and Section \ref{relatedwork} discusses related work. 
Section \ref{conclusion} concludes the paper.

\section{Background and Motivation}
\label{background}
In this section, we introduce fundamental concepts of stateful SPEs and dynamic scaling, discuss existing on-the-fly scaling methods and analyze the major overheads. 

\subsection{Stateful Stream Processing}
Computational tasks are usually represented as directed acyclic graphs (DAGs) in SPEs, where nodes denote operators each containing one or multiple processing instances, and edges represent inter-operator streams of data records.
Stateful operators, in particular, maintain internal states across distributed instances as disjoint partitions\cite{carbone2017state},
with routing tables in predecessors tracking this partitioning to ensure records correctly distributed to instances.

Due to the distributed nature of state maintenance, 
contemporary SPEs typically rely on checkpoints to ensure scaling correctness, leveraging their inherent consistency and treating scaling as a checkpoint-based restart operation.
However, usually scaling is needed only for specific bottleneck operators. 
Consequently, the overhead of creating global checkpoints with system-wide halts introduces significant but unnecessary latency during scaling\cite{carbone2015lightweight,heise2020aligned}. 
This defect motivates developing on-the-fly scaling methods that allow scaling operations on a subset of system components without necessitating global halts.

\subsection{On-the-fly Scaling}
\label{onthefly}
\begin{figure}[tb]
  \centering
    \includegraphics[width=0.9\linewidth]{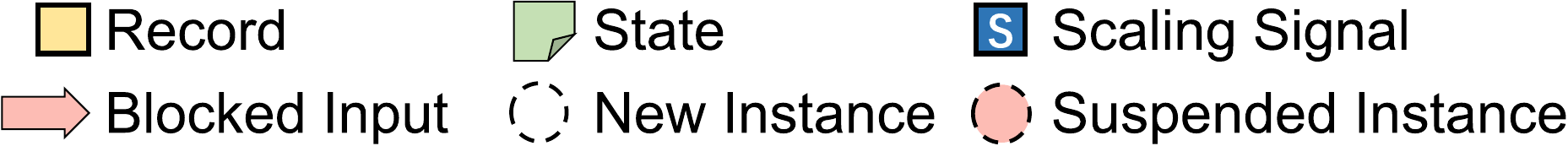}%
    
    \centering
    \begin{minipage}[b][5cm][b]{0.5\linewidth}
        \centering
        \begin{subfigure}[b]{0.75\linewidth}
            \centering
            \includegraphics[width=\linewidth]{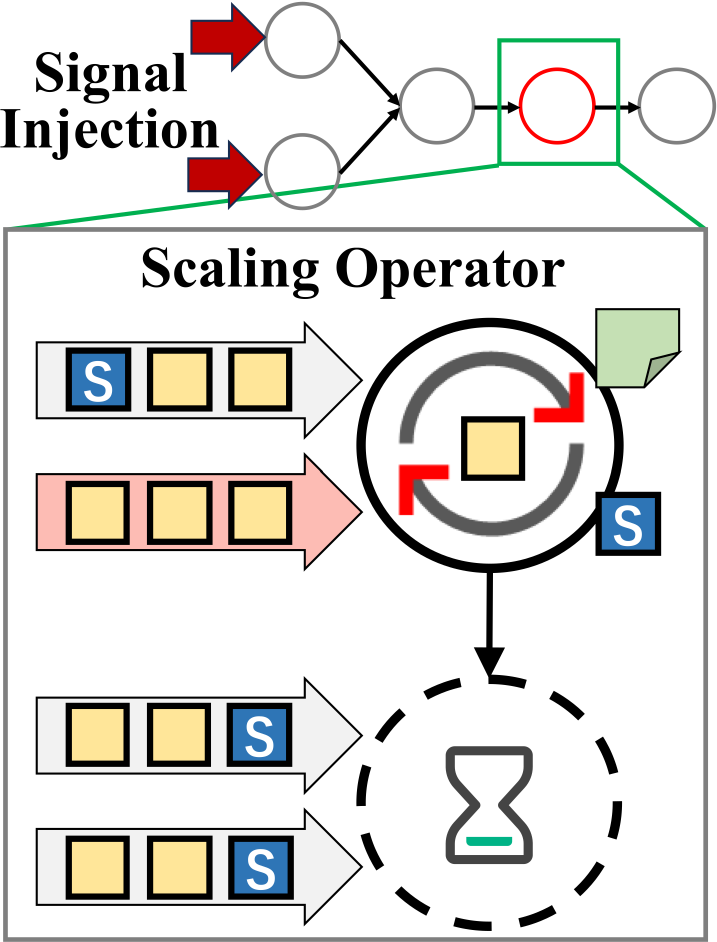}
            \caption{sync. phase}
            \label{generalized:sync}
        \end{subfigure}
    \end{minipage}
    \hfill
    \begin{minipage}[b][5cm][b]{0.48\linewidth}
        \centering
        \begin{minipage}[t]{0.85\linewidth}
            \centering
            \begin{subfigure}[t]{\linewidth}
                \centering
                \includegraphics[width=\linewidth]{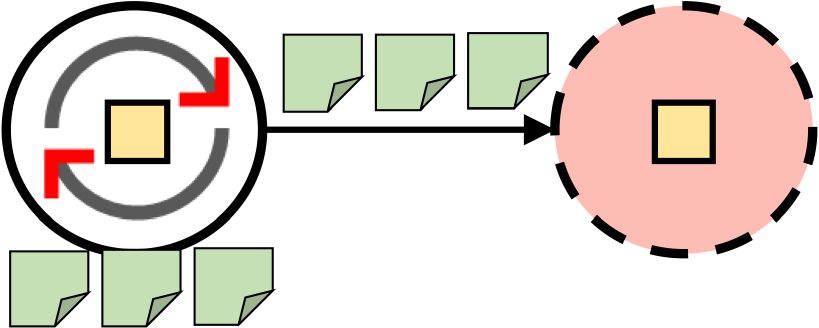}
                \caption{all-at-once migration}
                \label{generalized:once}
            \end{subfigure}
        \end{minipage}
        \centering
        \begin{minipage}[b][6.5\baselineskip]{0.85\linewidth}
            \centering
            \begin{subfigure}[b]{\linewidth}
                \centering
                \includegraphics[width=\linewidth]{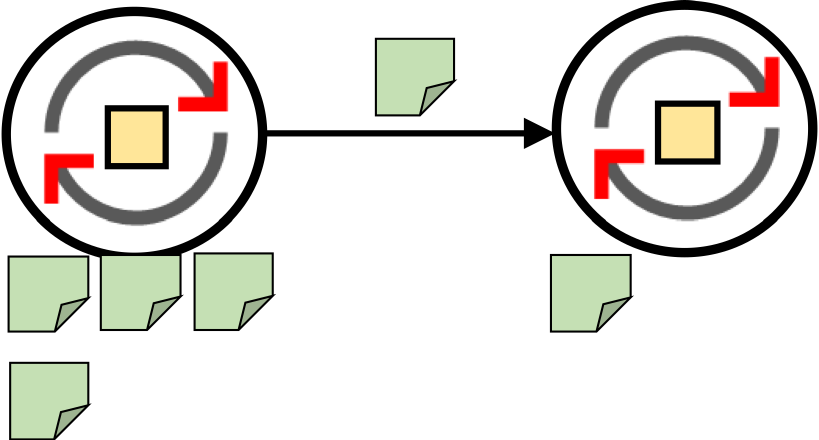}
                \caption{fluid migration}
                \label{generalized:fluid}
            \end{subfigure}
        \end{minipage}
    \end{minipage}
  \caption{
  {\color{black}
  Illustration of on-the-fly scaling (OTFS).}
  }
  \label{generalized}
\end{figure}

On-the-fly scaling (OTFS) in SPEs requires to perform scaling operations while maintaining continuous processing of incoming data streams.
Here we explain how such scaling mechanisms are designed. 
For clarity, we denote the operator to be scaled as the \textbf{scaling operator}, with its immediate predecessors and successors in the DAG termed \textbf{predecessors} and \textbf{successors}, respectively.
%
{\color{black}
We summarize existing OTFS methods and set up a generalized OTFS framework, comprising two key phases: synchronization\cite{gu2022meces,del2020rhino,mao2021trisk,apache_flink_elastic_scaling,rajadurai2018gloss} and state migration\cite{hoffmann2019megaphone,del2020rhino,wang2019elasticutor,gu2022meces}. With the commonality of the state of the arts, this framework provides a clear conceptual foundation for following discussions.
As shown in Fig.~\ref{generalized}, a simple example illustrates the generalized OTFS framework,} where synchronization is first achieved, followed by state migration from the original instance to a new one.

As depicted in Fig.~\ref{generalized:sync}, the synchronization phase begins with the source operator injecting the scaling signal, which propagates through the 
topology similar to a checkpoint barrier\cite{carbone2017state,akidau2015dataflow}.
While non-participating operators simply forward the signal after alignment, the predecessor operators must update their routing tables 
before signal propagation. 
Upon receiving the scaling signal, the original instance blocks the corresponding input channel to complete alignment. This alignment ensures that all the predecessors have updated their routing tables, guaranteeing that the states eligible for migration are no longer required for processing subsequent inputs, and can be safely transferred to the new instance once it completes runtime initialization and becomes fully operational.

In the state migration phase, {\color{black}there are two options.} Traditional approaches often adopt ``all-at-once migration" (Fig.~\ref{generalized:once}), wherein the original instance transfers all required states to the new instance in a single synchronized batch \cite{gulisano2012streamcloud,rajadurai2018gloss}.
{\color{black} Recent research has introduced ``fluid migration" that transfers states in a streaming manner \cite{hoffmann2019megaphone} (Fig.~\ref{generalized:fluid}). It allows each state to resume processing immediately upon arrival, rather than awaiting all remaining states to be transferred, thereby reducing overall processing suspensions.

However, the generalized OTFS framework with fluid migration} still encounters three challenges. 
First, lengthy signal propagation paths and alignment requirements cause 
significant propagation delays. 
%
Second, {\color{black} the system must passively wait for the associated states to be transferred, rather than proactively eliminating processing suspensions. 
Moreover, if the record at the head of the input channel is associated with the state unit at the tail of the migration sequence, fluid migration becomes as inefficient as all-at-once migration.}
%
Third, {\color{black} although fluid migration relaxes the global dependency to a per-unit sequential dependency, the latter still constrains scaling flexibility and introduce redundant queuing delays,} leading to extra processing latency and {\color{black} prolonging the scaling process.}

{\color{black}
Although some state-of-the-art methods have attempted to address these challenges, they still exhibit notable limitations.
Megaphone\cite{hoffmann2019megaphone} cannot achieve low propagation delays without non-universal SPE features\cite{murray2013naiad} (i.e., the separation of control and data planes and the frontier mechanism) or additionally introduced operators.
Similarly, Meces\cite{gu2022meces} incurs overhead from repeated back-and-forth migration of hot states and fails to preserve execution semantics under its fetch-on-demand method.
}

{\color{black}
Based on the observations and analysis of the common limitations in existing methods, we find the efficiency of on-the-fly scaling is constrained by three core factors:
\begin{itemize}
  \item \textbf{Propagation delays ($L_p$)}: the time required for scaling signals to propagate and synchronize across distributed instances.
  \item \textbf{Suspension delays ($L_s$)}: the cumulative duration during which record processing is blocked because of waiting for state migration.
  \item \textbf{Dependency delays ($L_d$)}: the overhead arising from interdependencies among migration units.
\end{itemize}
In other words, the overall on-the-fly scaling overhead can be expressed as: $L = L_p + L_s + L_d + L_o$, where $L_o$ represents inherent overheads,
such as physical resource initialization, state extraction, state serialization, etc., which are difficult to optimize through scaling mechanism design.

To validate this hypothesis, we design an ``extreme'' scaling solution called Unbound that focuses solely on performance without correctness.
Unbound updates routing tables and triggers state migration independently, thereby eliminating the need to propagate scaling signals. 
It converts record keys into ``universal keys'' that allow every local state to process any input record, eliminating processing suspensions due to incomplete state migration.
Although dependency still exists, the absence of processing suspensions prevents it from manifesting as increased latency. 
In this way, Unbound eliminates $L_p$ and $L_s$ and bypasses the impact of $L_d$ on latency. 
\begin{figure}[tb]
    \centering
    \includegraphics[width=\linewidth]{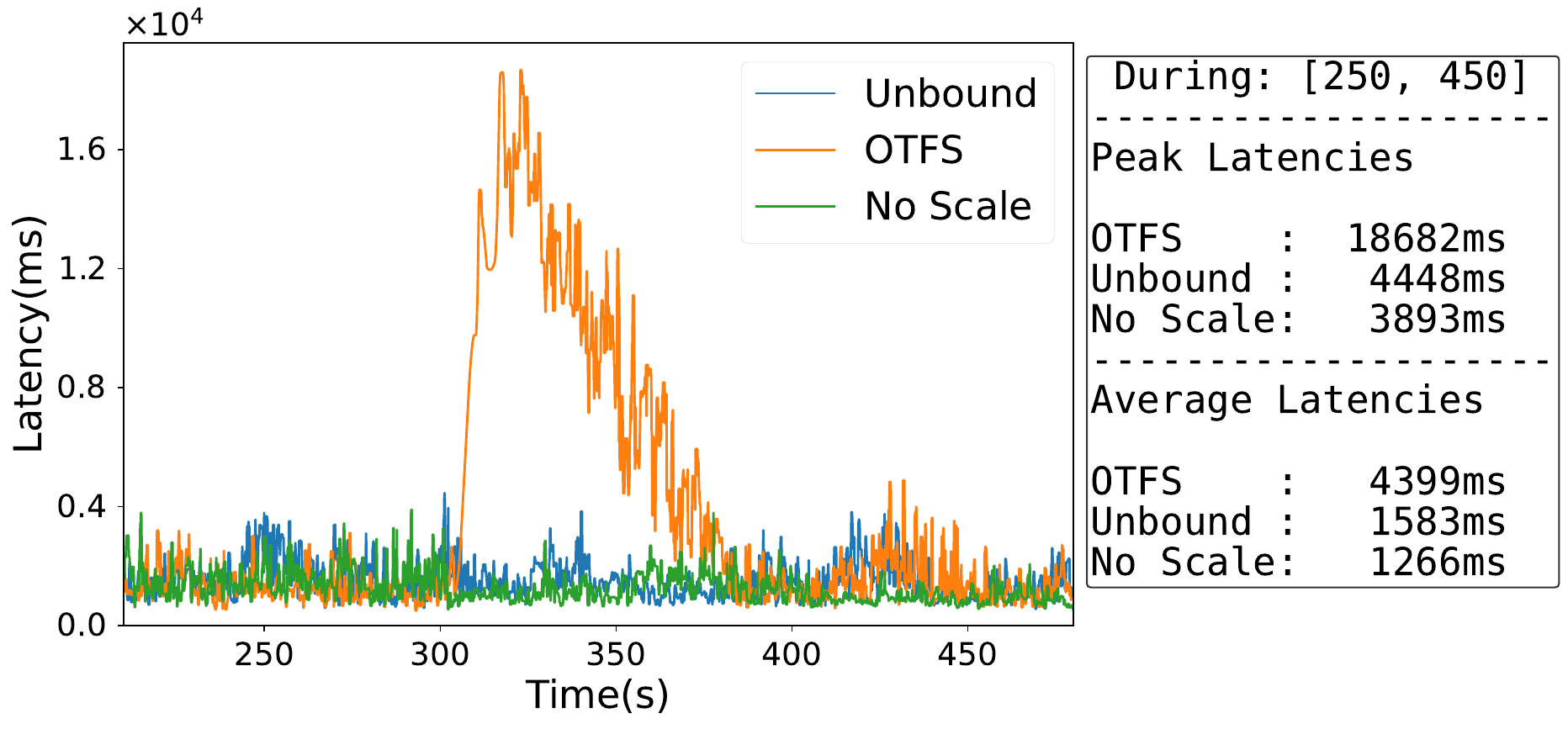}
    \caption{\color{black}
    The latency over time for: Unbound, OTFS (generalized on-the-fly scaling with fluid migration), and No Scale, measured using the experimental settings in Section V-A with the Twitch workload under a fixed input rate.
    } 
    \label{fig:preliminary}
\end{figure}
The experimental results in Fig.~\ref{fig:preliminary} demonstrate that, compared to OTFS, Unbound reduces the average latency from 3.47x to 1.25x relative to No Scale, and the peak latency from 4.8x to 1.14x. 
This indicates that the performance difference between Unbound and No Scale is no longer significant,
confirming our hypothesis that the aforementioned three issues are the core factors in on-the-fly scaling.}

\section{System Design}
\label{design}
DRRS introduces three novel mechanisms to 
{\color{black}mitigate}
scaling-induced performance degradation in stateful SPEs:

\begin{itemize} 
{\color{black}\item \textbf{Decoupling and Re-routing} eliminates the propagation delays of scaling signals 
through decoupling them and directly injecting them into predecessors. 
\item \textbf{Record Scheduling} reduces processing suspensions during state migration with semantic-preserving adjustments to the internal execution order of records. 
\item \textbf{Subscale Division} alleviates dependency-related overhead between state units 
by partitioning the scaling process into isolated subscales.}
\end{itemize}

\begin{figure}[tb]
    \centering
    \includegraphics[width=0.8\linewidth]{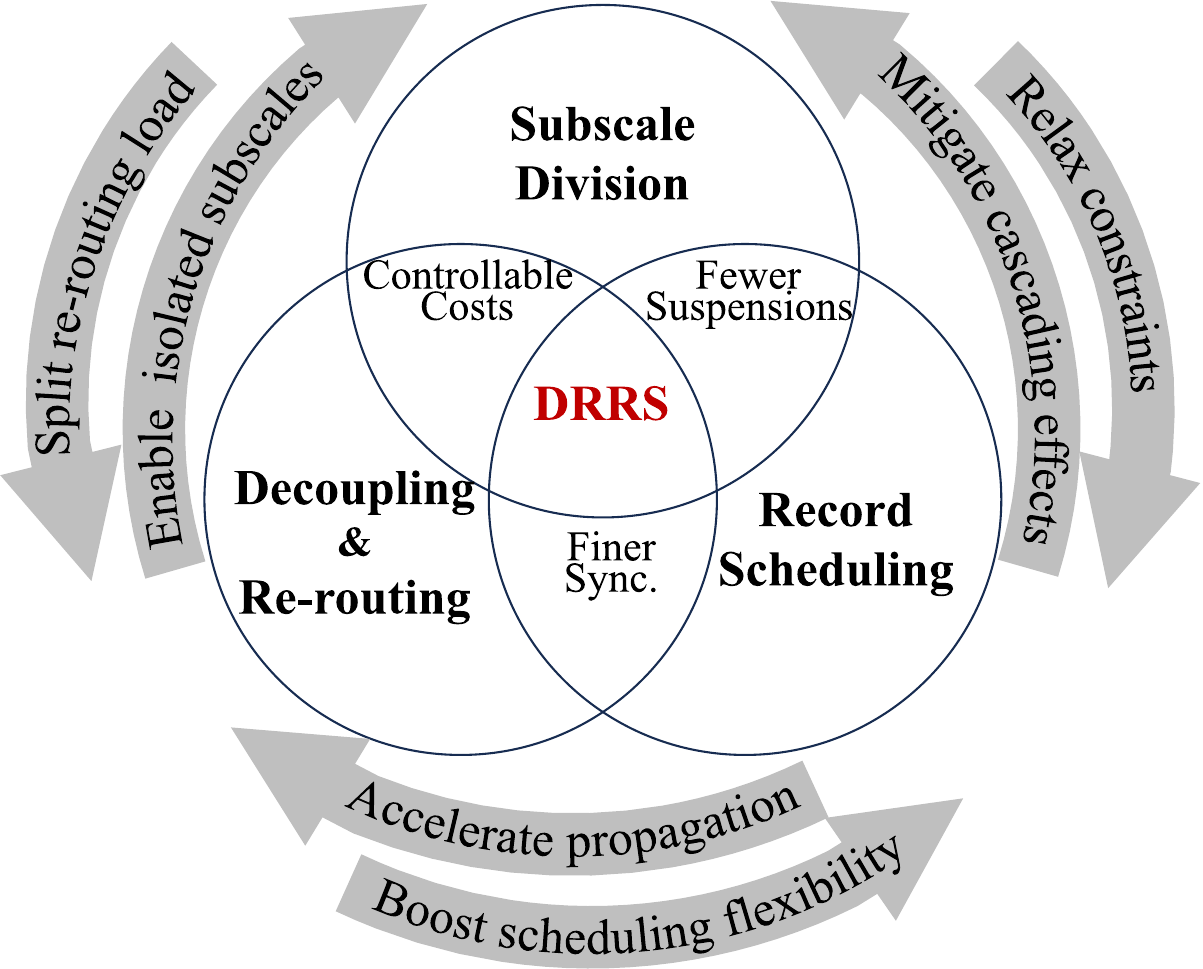}
  \caption{\color{black}Synergy between DRRS Mechanisms.}
  \label{fig:connections}
\end{figure}

Moreover, these mechanisms interact synergistically to reinforce one another, as illustrated in Fig.~\ref{fig:connections}.
First, Decoupling and Re-routing enables earlier state migration, providing additional flexibility for Record Scheduling.
This reduces processing suspensions, which in turn mitigates the propagation blockages.
Second, Decoupling and Re-routing serves as a foundation for Subscale Division by preventing signal propagation interference between subscales. 
Conversely, Subscale Division reduces Re-routing overhead by limiting operations to fewer state units, improving resource utilization, especially in network buffers.
Finally, while subscales operate independently, suspensions in one instance will inevitably impact all related subscales.
This cascading effect is mitigated by Record Scheduling with reduced processing suspensions. 
Moreover, the fine-grained partitioning in Subscale Division alleviates scheduling restrictions, further minimizing suspensions.

The rest of this section presents the details of the three mechanisms as the core design of DRRS.

\subsection{Decoupling and Re-routing}
\label{decoupingrouting}
{\color{black}
Existing techniques~\cite{gu2022meces,del2020rhino,apache_flink_elastic_scaling} for scaling signal propagation in stateful SPEs predominantly rely on mechanisms derived from other synchronization signals, such as checkpoint barriers.}
However, this design presents two significant limitations when applied to on-the-fly scaling.
First, on-the-fly scaling does not require global consistency for state migration. As a result, the source-injection and mandatory alignment mechanisms designed to guarantee global consistency become superfluous, introducing unnecessary synchronization overhead.
Second, even with a direct predecessor-injection approach, the barrier's dual functionality, i. e., serving simultaneously as routing confirmation and migration trigger, introduces an inherent conflict.
On one hand, enhanced adaptability to workload changes demands early triggering of state migration, implying that in-flight data should be bypassed for immediate effect.
On the other hand, routing confirmation must adhere to the orderly dataflow to ensure the correct coordination of routing table updates.

\newcommand{\figheighttwo}{2.4cm}
\newcommand{\figwidthtwo}{0.19\linewidth}

\begin{figure*}[t]
  \centering
  \begin{subfigure}[t]{\linewidth}
    \centering
    \includegraphics[width=0.85\linewidth]{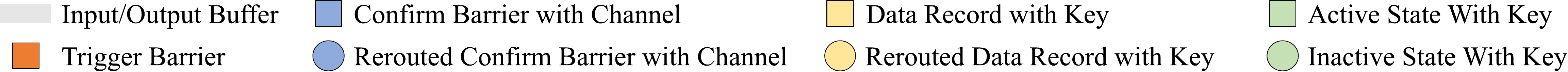}%
  \end{subfigure}

  \begin{subfigure}[t]{\figwidthtwo}
    \centering
      \includegraphics[height=\figheighttwo]{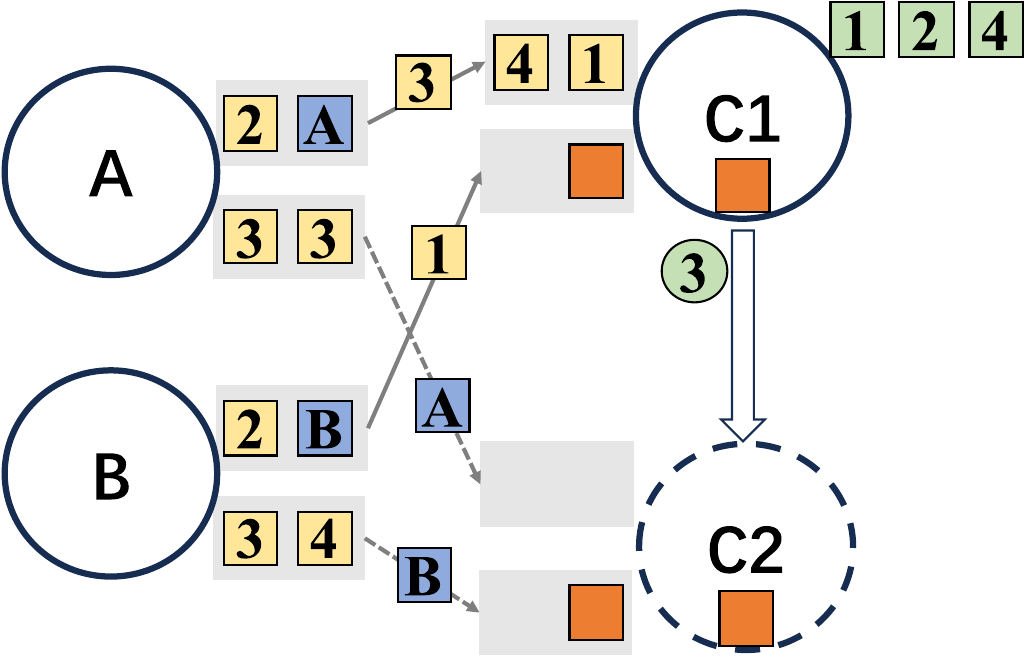}%
    \caption{}
    \label{dr:a}
  \end{subfigure}
  \hfill
  \begin{subfigure}[t]{\figwidthtwo}
    \centering
      \includegraphics[height=\figheighttwo]{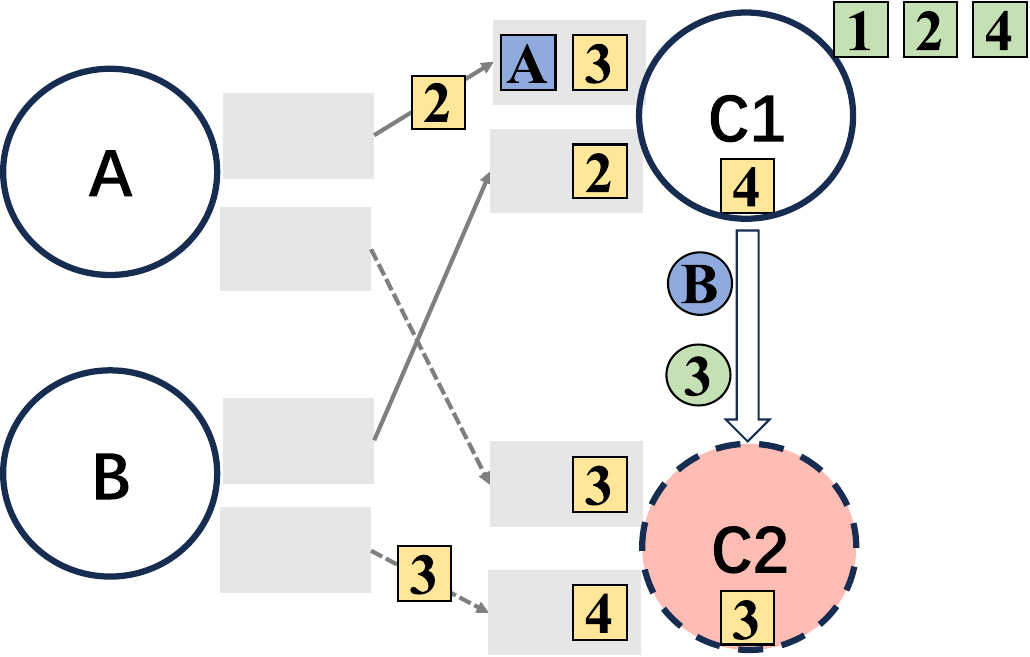}%
    \caption{}
    \label{dr:b}
  \end{subfigure}
  \hfill
  \begin{subfigure}[t]{\figwidthtwo}
    \centering
      \includegraphics[height=\figheighttwo]{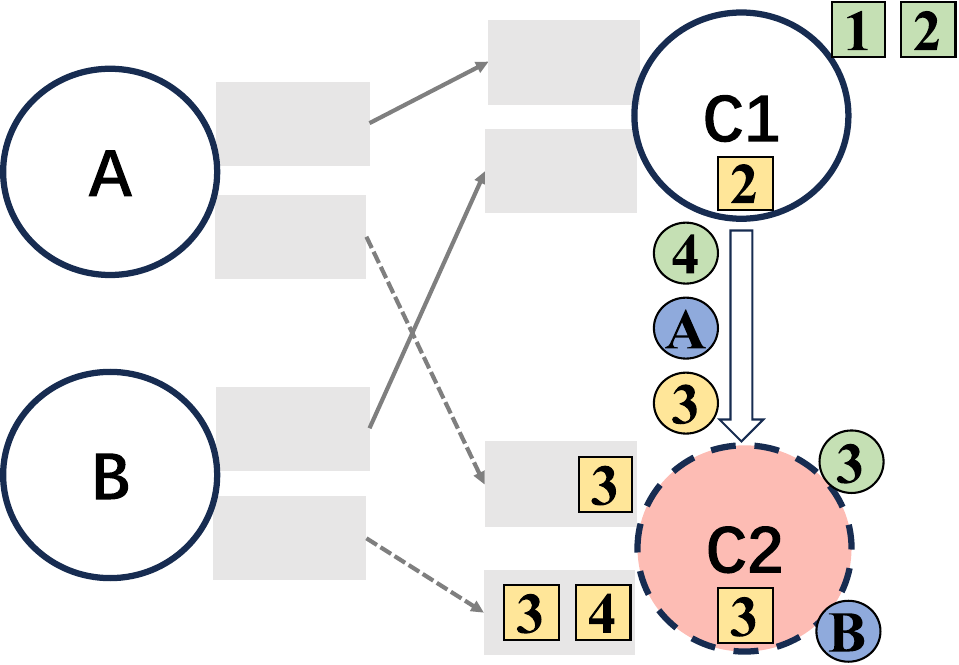}%
    \caption{}
    \label{dr:c}
  \end{subfigure}
  \hfill
  \begin{subfigure}[t]{\figwidthtwo}
    \centering
      \includegraphics[height=\figheighttwo]{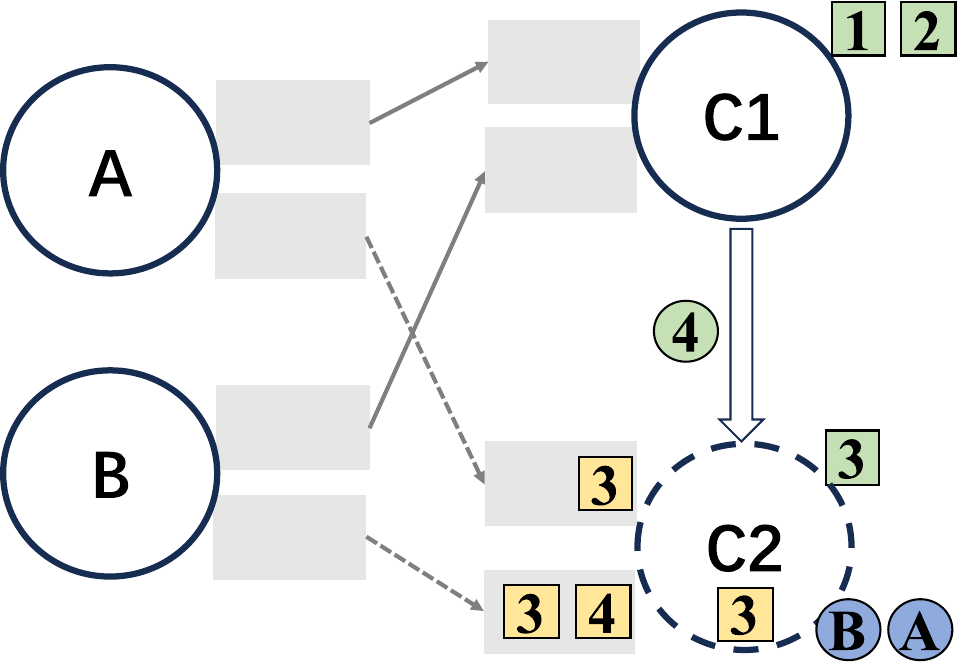}%
    \caption{}
    \label{dr:d}
  \end{subfigure}
  \hfill
  \begin{subfigure}[t]{\figwidthtwo}
    \centering
      \includegraphics[height=\figheighttwo]{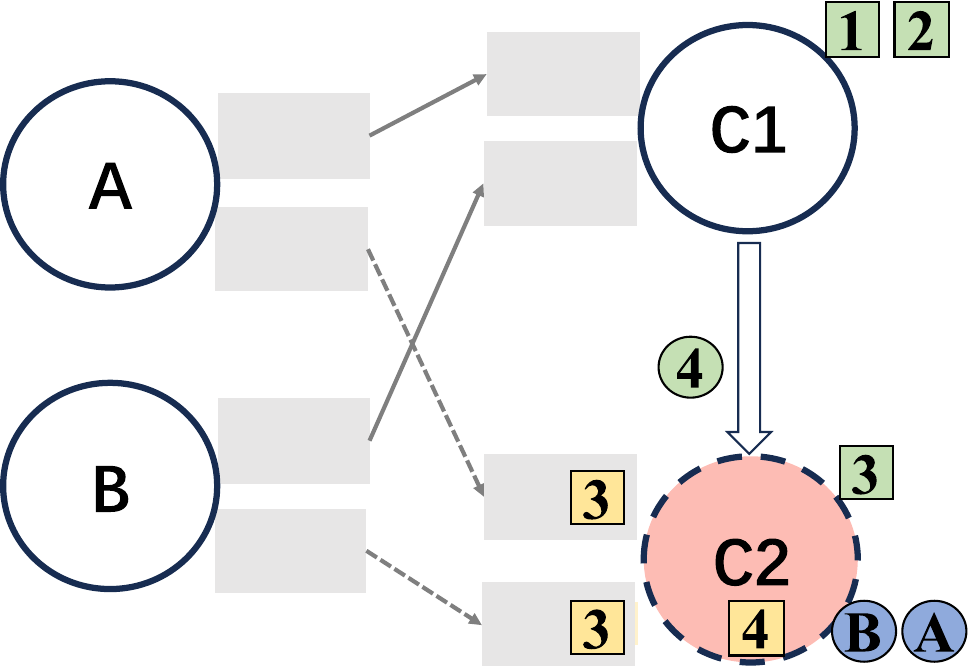}%
    \caption{}
    \label{dr:e}
  \end{subfigure}
  \caption{Illustration of the Decoupling and Re-routing mechanism.
  } 
  \label{fig:dr}
\end{figure*}

Based on the above observations, we propose \textbf{Decoupling and Re-routing}, a fine-grained synchronization mechanism that not only shifts from source-injection to predecessor-injection, but also decomposes the conventional dual-purpose barrier into two distinct signals:
\begin{itemize}
    \item \textbf{Trigger barrier}: a priority message that bypasses all in-flight data (both input and output caches) to expedite state migration initiation.
    \item \textbf{Confirm barrier}: 
    {\color{black}mostly a non-priority message to ensure ordered propagation,} but treated as a priority message only in the output cache with any bypassed records subsequently re-queued as needed to guarantee correct record routing,
    and {\color{black} reverted to non-priority upon arriving at the scaling operator.}
\end{itemize}
%
%

The Decoupling and Re-routing mechanism sequentially injects trigger and confirm barriers at the predecessor operators, and propagates them 
according to their respective patterns.
{\color{black} Upon receiving any trigger barrier, scaling instances immediately initiate state migration and ignore any subsequent trigger barriers.
For confirm barriers, although alignment is still required, 
the Decoupling and Re-routing mechanism shifts the alignment process from the senders to the receivers of the migrated state by re-routing all confirm barriers.
This transforms conventional explicit alignment into an ``implicit alignment" without input blocking.
Additionally, because migration commences prior to the alignment of confirm barriers, re-routing is also responsible for handling records associated with the states that have already migrated out to preserve semantics.}

Fig.~\ref{fig:dr} illustrates how the stateful operator $C$ scales from one to two instances, involving the addition of instance $C_2$ and the state migration of keys 3 and 4 from $C_1$ to $C_2$. 
For simplicity, we focus solely on the sub-DAG and records involved in the scaling process, denoting the state and record with key $i$ as $S_i$ and $R_i$, respectively.
As shown in Fig.~\ref{dr:a}, upon receiving a scaling signal injection command, 
predecessor instances A and B update their routing tables and broadcast both trigger barriers and confirm barriers. 
%
After the emission of those signals, any $R_3$ and $R_4$ records in the output cache originally targeted for $C_1$ but bypassed by the confirm barrier must be redirected to the newly-created output cache targeting $C_2$, while preserving their relative ordering.
During state migration, as depicted in Fig.~\ref{dr:b}, $C_1$ continues processing $R_1$ and $R_2$ (associated with non-migrating states), as well as $R_4$ (associated with $S_4$ still awaiting migration).
Meanwhile, $C_2$ suspends processing when encountering 
its first input record $R_3$.
This suspension arises not only because $S_3$ is still migrating, but also to ensure implicit alignment, which preserves execution semantics.
Consequently, 
simply receiving $S_3$ is insufficient to lift $C_2$'s suspension, since rerouted confirm barrier $B$ has not yet arrived.
In addition to re-routing confirm barriers, $C_1$ must also re-route records associated with the states that have migrated out, such as $R_3$ in Fig.~\ref{dr:c}.
These rerouted records are handled as special events and are not affected by processing suspension.

After achieving implicit alignment, as illustrated in Fig.~\ref{dr:d}, state $S_3$ can 
transfer from inactive to active and enable $C_2$ to resume processing record $R_3$.
Subsequently, $C_2$ encounters another suspension while processing $R_4$.
This new suspension arises solely because $S_4$ is not locally available as shown in Fig.~\ref{dr:e}, and processing resumes immediately once $S_4$ arrives,
indicating the completion of the entire scaling process in operator $C$.

\begin{figure}[tb]
    \centering
    \includegraphics[width=\linewidth]{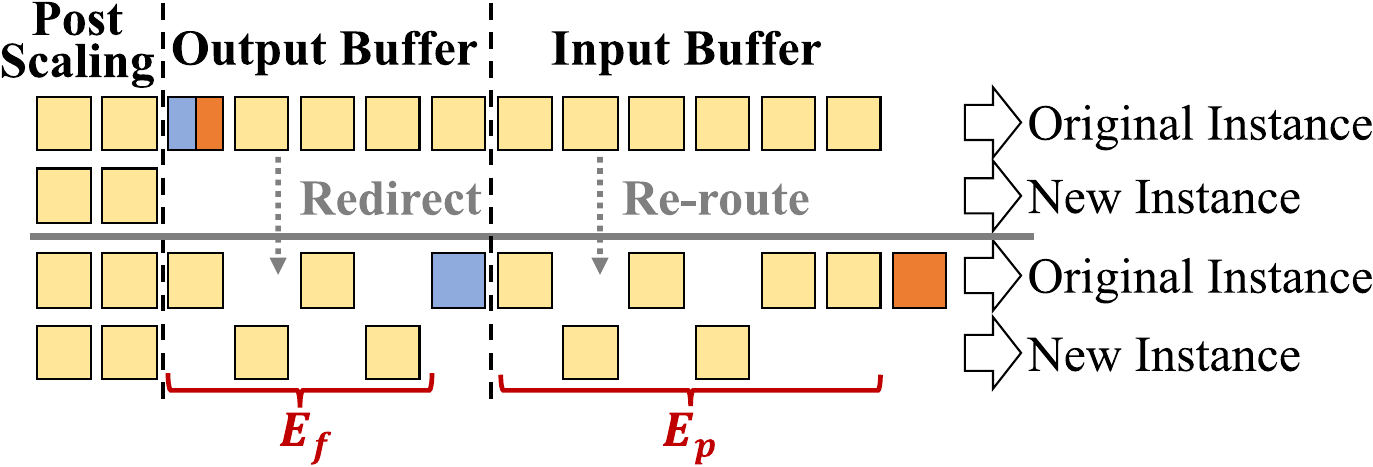}
    \caption{\color{black}
    Illustration of streams under Coupled vs. Decoupled Signal Synchronization. The top half depicts synchronization with a conventional coupled signal, while the bottom demonstrates decoupling and re-routing, organizing affected records into two epochs ($E_f$ and $E_p$).
    }
    \label{fig:implicat_alignment}
\end{figure}

{\color{black}
In addition to the above example in Fig.~\ref{fig:dr} focused on how Decoupling and Re-routing reduces propagation delays, we next elaborate how the mechanism ensures preserved execution semantics in scaling.
As shown in Fig.~\ref{fig:implicat_alignment}, 
for a scaling instance's input, 
conventionally the predecessors only forward records 
after signaling routing confirmation.
Therefore, as shown in the top half, the input stream for the new instance is not generated until the coupled signals are broadcast.
In contrast, as shown in the bottom half, Decoupling and Re-routing allows the new stream to be generated earlier, thereby enabling sooner processing in the new instance.
All pending records can be divided by the confirm barrier into two epochs: $E_p$ (preceding epoch) and $E_f$ (following epoch).
A new instance switches from $E_p$ to $E_f$ once the implicit alignment is achieved, i.e., all the rerouted confirm barriers arrive. 
The $E_f$ records are redirected to the output cache 
to align with the updated routing table, 
thus 
naturally preserving execution semantics.
The $E_p$ records 
may have their related states already migrated out by the time they are processed by the original instance, and 
consequently need to be re-routed to the new instance for processing. 
Since the re-routing mechanism itself is ordered, execution semantics are preserved for these records as well. 
%
}

It is worth noting that Decoupling and Re-routing is highly {\color{black}compatible with other mechanisms that also rely on messages.
Affected data-driven messages (e.g., triggers, watermarks) are simply duplicated to both input streams---those of the original and the new instance---during redirection and re-routing, ensuring correct propagation in both.}
Since rerouted records are sourced exclusively from the scaling instance's input cache (in $E_p$), the associated overhead remains bounded by the cache size.

\subsection{Record Scheduling}

\begin{figure}[tb]
  \begin{minipage}{\linewidth}
    \centering
    \includegraphics[width=0.7\linewidth]{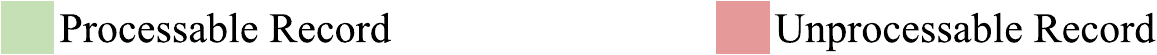}
  \end{minipage}
\begin{minipage}{\linewidth}
    \begin{subfigure}[t]{0.49\linewidth}
      \centering
      \includegraphics[width=\linewidth]{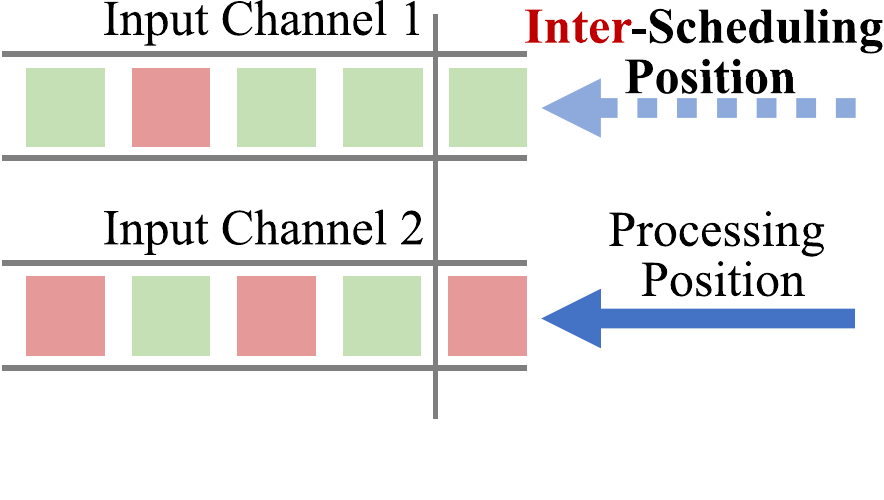}
      \caption{Inter-Channel Scheduling.} 
      \label{reorder:inter}
    \end{subfigure}
    \hfill 
    \begin{subfigure}[t]{0.49\linewidth}
      \centering
      \includegraphics[width=\linewidth]{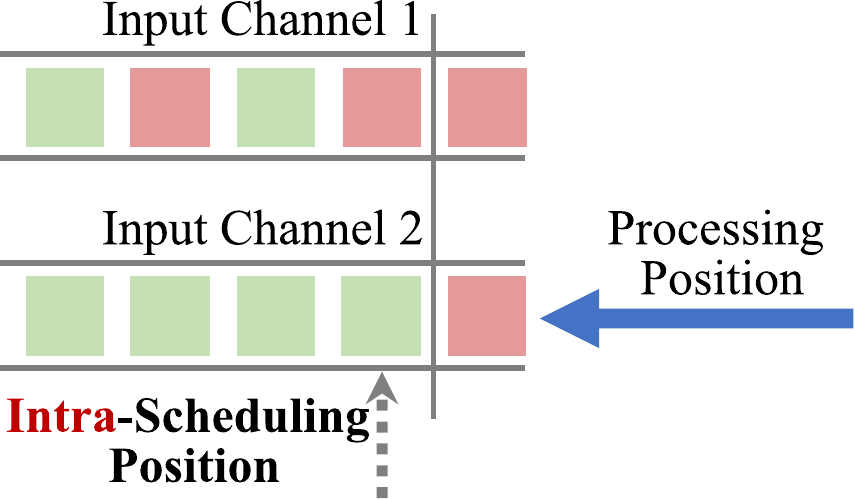}
      \caption{Intra-Channel Scheduling.}
      \label{reorder:intra}
    \end{subfigure}
  \end{minipage}
  \caption{Reducing suspensions with {\color{black}Record Scheduling.}}
  \label{fig:reorder}
\end{figure}

Although fluid migration reduces processing suspension time compared to conventional all-at-once migration, {\color{black} it still lacks a proactive mechanism for handling suspensions, 
and potentially causes optimization fallbacks.} 
To address this issue, we begin with 
examining the status of input channels in suspended instances and define a channel 
as \textit{processable} if its frontmost record can be processed. As shown in Fig.~\ref{fig:reorder}, when an instance suspends because its currently active channel is not processable, other channels may still remain processable, and even the suspended channel itself may contain processable records in later positions.
Motivated by these observations, we propose the 
{\color{black} \textbf{Record Scheduling} }mechanism, which prevents processing suspensions through semantic-preserving adjustments to {\color{black} engine-level record order.
While certain operators can modify record order at the operator-level, we first clarify the fundamental distinction between two ordering concepts in streaming systems literature: 
\begin{itemize}
    \item Engine-level order (execution order): The sequence in which records are scheduled and delivered to operators at the SPE level.
    \item Operator-level order (logical order): The sequence in which records are processed by operators, often determined by the operator's specific semantics or processing logic.
\end{itemize}

Existing methods \cite{traub2018scotty,li2008out} for record ordering primarily address out-of-order processing based on watermarks. However, these approaches are not suitable for preventing suspension, as they are confined to reorganizing records that have already been delivered by the engine. They merely wait for the engine to resume delivery once suspension ends.
In contrast, the Record Scheduling mechanism can proactively prevent processing stalls at the engine level,} by switching to processable channels (\textbf{Inter-channel Scheduling}) or temporarily bypassing unprocessable records within the same channel under specific constraints (\textbf{Intra-channel Scheduling}),
thereby addressing inefficiencies while {\color{black} ensuring the record delivery order remain consistent with operator semantics.}


Inter-channel 
{\color{black} Scheduling}
allows instances to dynamically switch to alternative processable channels when encountering unprocessable records in its current processing channel.
As shown in Fig.~\ref{reorder:inter}, if processing is suspended in channel 2, the system can proactively switch to processable channel 1 to continue operation, rather than suspending and waiting for the state migration.
This cross-channel order adjustment does not 
violate execution semantics, as the execution order across channels is inherently non-deterministic due to factors like network latency.
Furthermore, Inter-Channel 
{\color{black} Scheduling}
works in concert with the Decoupling and Re-Routing 
{\color{black} 
to enable finer-grained scaling synchronization, which can be viewed as ``fluid confirmation".}
Benefiting from dynamic channel adjustment, each channel can switch epochs independently without waiting for all rerouted confirm barriers to arrive, completely eliminating alignment requirements.
This cooperation further reduces processing suspensions, especially in the scenarios with asymmetric data distribution across channels.

A limitation of Inter-channel 
{\color{black} Scheduling}
is that it may fail to prevent suspensions in instances with few channels.
To address this limitation, we introduce Intra-channel 
{\color{black} Scheduling}, which enables processing order adjustments within individual channels. As shown in Fig.~\ref{reorder:intra}, when all the channels are unprocessable, the instance bypasses the foremost unprocessable record in channel 2 and proceeds with subsequent processable records. 
However, unlike its inter-channel counterpart, Intra-channel Scheduling faces additional constraints on execution order adjustments due to the presence of data-driven messages in modern SPEs.
{\color{black} While these messages are processed at the operator level, their propagation introduces implicit dependencies between records at the engine level, further complicating execution order preservation.}
Typically, under Event Time semantics, scheduling must avoid crossing time-semantics signals such as watermarks or global window triggers, as these signals are critical for processing result consistency. 
In other task scenarios, scheduling principles can be tailored to specific requirements. For example, tasks that tolerate approximate results or where the processing order does not affect the final outcome can relax ordering constraints, such that only preserving exactly-once semantics is sufficient, enabling flexible Intra-channel Scheduling.


\subsection{Subscale Division}
\label{sec:subscale_division}
In addition to processing suspensions, state migration also encounters challenges like inter-dependencies and coordination between state units. 
These factors introduce additional costs and limitations on flexibility, referred to as \textbf{dependency-related overhead} in this work. 
These constraints, such as the global dependency in all-at-once migration and serial dependency in fluid migration, become particularly significant when scaling involves a large number of state migrations.

\begin{figure}[tb]
  \centering
  \begin{subfigure}[t]{0.49\linewidth}
    \centering
      \includegraphics[height=2.4cm]{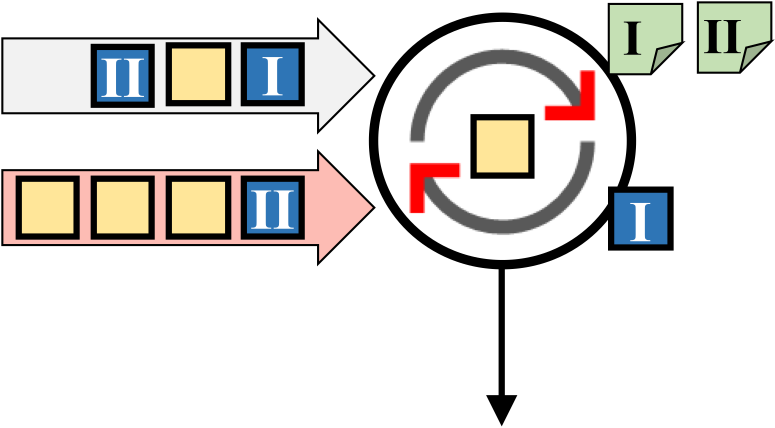}%
    \caption{Naive Division}
    \label{subscale:a}
  \end{subfigure}
  \hfill
  \begin{subfigure}[t]{0.49\linewidth}
    \centering
    \includegraphics[height=2.4cm]{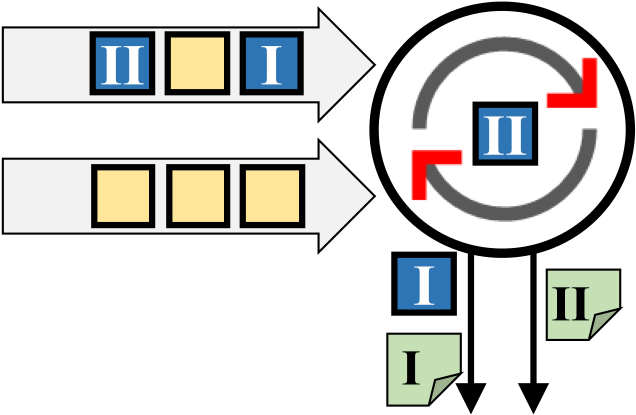}%
    \caption{Subscale Division}
    \label{subscale:b}
  \end{subfigure}
  \caption{Naive Division vs. Subscale Division.}
  \label{fig:subscale}
\end{figure}

To fundamentally reduce dependency-related overhead, we can partition the migrating states into multiple independently migrated subsets, and thus reduce the number of state units involved in any single dependency relationship.
A Naive approach can assign each subset to a separate scaling operation, decomposing a ``large" scaling process into multiple ``small" scaling operations. 
However, the Naive Division presents two critical problems:

\subsubsection{Non-migration Overhead Accumulation of each scaling operation}
Each scaling operation introduces additional overhead beyond migration-related costs, including resource initialization and synchronization.

\subsubsection{Interference between scaling operations}
As depicted in Fig.~\ref{subscale:a}, once the instance receives scaling signal I, it must block the corresponding input channel to achieve alignment, which consequently impedes the propagation of synchronization signal II. 
This sequential constraint reveals that the naive approach merely transforms state unit dependencies into subset-level dependencies rather than fundamentally reducing them, significantly extending the overall scaling duration and compromising the system adaptability to workload variations.

To address the two problems, we propose \textbf{Subscale Division}, a novel mechanism built upon the Decoupling and Re-routing mechanism. 
Instead of assigning each subset of state units to separate scaling operations, Subscale Division introduces multiple subscales within a single scaling operation, significantly reducing the accumulated non-migration overhead. 
Furthermore, by leveraging Decoupling and Re-routing's non-blocking scaling signals, each subscale can operate independently without interference, avoiding the subset-level dependencies inherent in Naive Division and reducing synchronization overhead in each subscale.
Fig.~\ref{subscale:b} demonstrates how Subscale Division works.
The trigger and confirm barriers of different subscales can propagate simultaneously through the data flow without interference. 
When an instance receives a trigger barrier of a new subscale, it sets up an independent migration path.
Along these paths, each subscale independently migrates and re-routes its states, records, and confirm barriers, which enables flexible scheduling. 
Furthermore, as Subscale Division is fundamentally based on Decoupling and Re-routing, the guarantees regarding execution semantics are preserved.



Subscale Division significantly enhances scaling flexibility through fine-grained control and runtime-adaptive migration.
The independence of each subscale enables more flexible scheduling based on real-time system conditions, e.g., 
executing multiple subscales in parallel accelerates scaling during resource-abundant periods, while a more conservative sequential execution helps prevent contention in resource-constrained environments.

\section{Implementation}
\label{implementation}
In this section, we present the implementation of DRRS, including the system architecture and the intricate collaboration between the DRRS modules.
We also discuss concurrent DRRS executions, and DRRS's compatibility with fault tolerance as a critical feature in modern SPEs.

\subsection{System Architecture}
\label{sec:architecture}

\begin{figure}[tb]
  \centering
  \includegraphics[width=\linewidth]{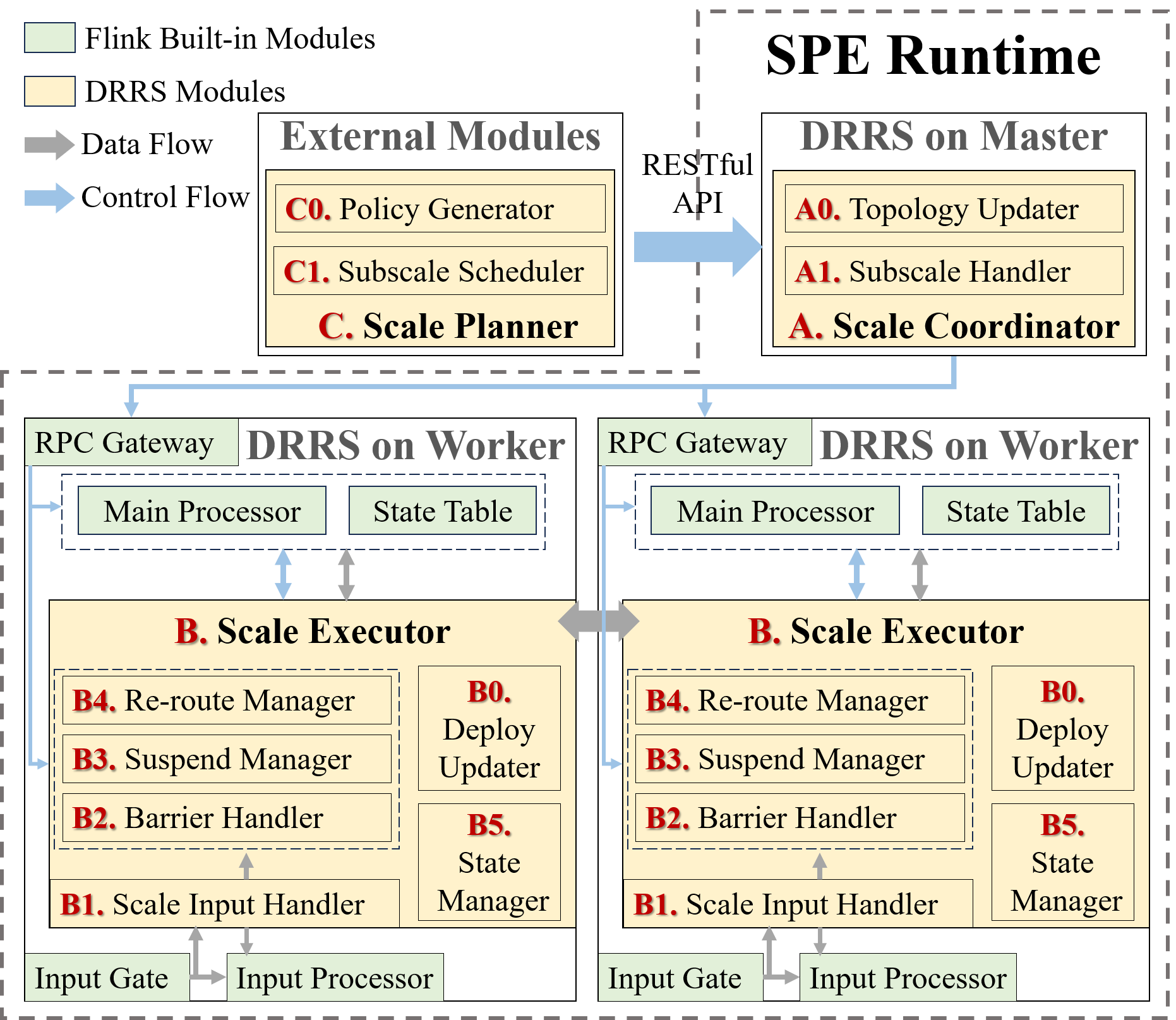}
  \caption{DRRS System Architecture.}
  \label{fig:drrs}
\end{figure}

We implement DRRS as a plugin for Apache Flink\cite{flink}, one of the most popular open-source engines in the field of stream computing.
As illustrated in Fig.~\ref{fig:drrs}, DRRS primarily consists of three main components: 
\begin{itemize}
\item \textbf{Scale Coordinator} {\color{black}(\textit{A})} runs on the master node, responding to scaling-related requests and coordinating multiple worker nodes.
\item \textbf{Scale Executor} {\color{black}(\textit{B})} runs on each worker node and carries out the physical operations triggered by \textit{A}.
\item {\color{black}\textbf{Scale Planner} (\textit{C}) generates scaling plans based on user-defined policies, including triggering conditions, state partitioning strategies, and subscale scheduling.
Since Scale Planner is a necessary but orthogonal component to our focus,
we have designed it as an independent module that communicates with \textit{A} to invoke scaling, allowing users to easily configure it.}
\end{itemize} 

In our default implementation of the Scale Planner, we adopted the following strategies:
{\color{black} 
Policy Generator (\textit{C0}) uses a user-request-based trigger that uses a uniform repartitioning strategy for state redistribution;
Subscale Scheduler (\textit{C1}) 
lexicographically divides states into subsets as equally sized as possible} and schedules them using a simple greedy strategy which prioritizes subscales migrating to instances with the fewest held keys, thus rapidly involving new instances in the computation.
To avoid potential resource contention, we also set a concurrency threshold, 
limiting each node to participating in no more than two subscale operations simultaneously.

The three core components of DRRS operate through coordinated interactions. 
Initially, the system updates deployments and initializes scaling-related modules.
{\color{black} The Topology Updater (\textit{A0}) generates update commands based on \textit{C0}'s scaling decisions and the Deploy Updater (\textit{B0}) adjusts physical resources and cluster deployments accordingly.}
Meanwhile, the Scale Input Handler {\color{black}(\textit{B1})} replaces Flink's native Input Handler to identify and process the records and signals essential for scaling.

Once these initial steps are completed, {\color{black}\textit{C1} begins sending subscale commands.} 
Upon receiving these commands, the Subscale Handler {\color{black}(\textit{A1})} commands the predecessor operators to initiate the scaling signal injection according to the Decoupling and Re-routing mechanism.
In scaling operators, {\color{black}\textit{B1}} identifies and routes the incoming records or signals to the appropriate modules for processing:
\begin{itemize}
\item {\color{black}Trigger and Confirm barriers are handled by Barrier Handler (\textit{B2}), the former initiating state migration and the latter being re-routed to relevant instances.}
\item Records ready for processing are handled by the native Input Handler, including those associated with the states that require no migration, have completed migration, or are still awaiting migration.
\item Temporarily unprocessable records are handled by the Suspend Manager {\color{black}(\textit{B3})}. Processing is suspended only when all swappable records are unprocessable, and the system waits until any of them becomes processable.
\item Records related to the states that have been migrated out are managed by the Re-route Manager {\color{black}(\textit{B4})}, which implements configurable capacity-based or timeout-based re-routing strategies. 
{\color{black}Additionally, the re-routing of confirm barriers causes an immediate re-route of records in network caches to maintain the relative order between records and barriers.}
\end{itemize}

Once {\color{black}\textit{C1}} notifies that the last subscale instruction has been sent, {\color{black}\textit{A}} and {\color{black}\textit{B}} performs cleanup and returns to a non-scaling status after all ongoing subscales are completed.
Since Flink lacks random input access, we implement a bounded buffer to cache up to 200 records, supporting the Record Scheduling mechanism for handling temporarily unprocessable records.
{\color{black}
Furthermore, DRRS remains inactive until explicitly activated by scaling requests. Upon the completion of scaling, the framework automatically releases all allocated resources, such as memory buffers and control threads. This ensures that no DRRS components (except for a few monitors and handlers) remain in runtime memory in regular operation, thereby preventing any potential interference with normal execution outside of scaling events.}

\subsection{Concurrent DRRS Executions}
{\color{black}
DRRS's fine-grained approach enables concurrent scaling executions, which is overlooked in previous research but crucial for system adaptability.}
By focusing on the scaling operator and the output/input components of its immediate predecessors/successors, DRRS limits potential interference to two specific scenarios:

\subsubsection{Concurrent scaling requests targeting the same operator, typically occurring during rapid load fluctuations} 
The system immediately terminates the preceding scaling operation and initiates a new one, avoiding redundant data migrations as the latter scaling supersedes the former.

\subsubsection{An operator serving both as a scaling operator and as a predecessor/successor to another scaling operator}
The system must guarantee consistent deployment updates, ensuring that new instances adopt the same input/output configurations as the existing ones. 


\subsection{Compatibility with Fault Tolerance}
\label{sec:compatibility}

\begin{figure}[tb]
  \centering
    \begin{subfigure}[t]{0.49\linewidth}
    \centering
      \includegraphics[height=2cm]{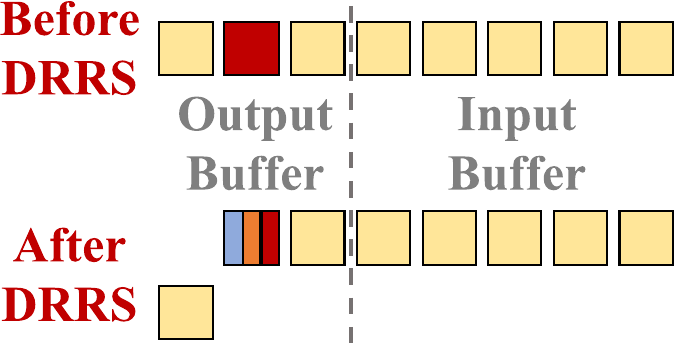}%
    \caption{In Output Buffer}
    \label{compatibility:a}
  \end{subfigure}
  \hfill
  \begin{subfigure}[t]{0.49\linewidth}
    \centering
    \includegraphics[height=2cm]{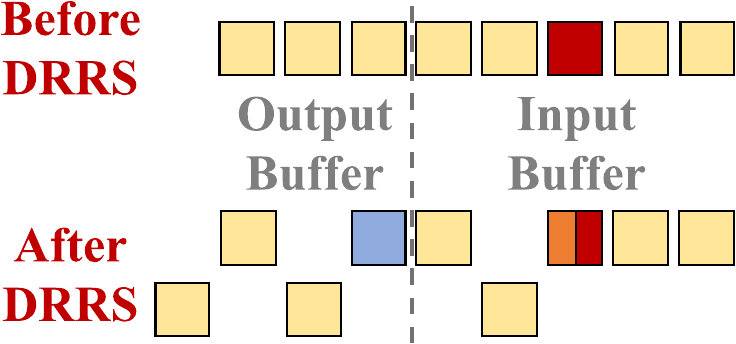}%
    \caption{In Input Buffer}
    \label{compatibility:b}
  \end{subfigure}

  \caption{\color{black} Impact of DRRS on checkpoint barrier transmission:  The red one represents the checkpoint barrier. The top/bottom half shows the stream sequence before/after DRRS signal injection.
  }
  \label{fig:compatibility}
\end{figure}


DRRS maintains seamless compatibility with SPE's fault tolerance features, by interacting with system-level control messages.
Similar to prior work \cite{hoffmann2019megaphone,gu2022meces}, DRRS prevents concurrent execution of fault tolerance mechanisms and scaling operations to avoid potential conflicts.
However, instead of delaying one operation until the other completes entirely, 
DRRS's focus on the scaling operator and its upstream neighbors provides an efficient alternative that synchronizes DRRS signals with checkpoint barriers.
{\color{black}
As illustrated in Fig.~\ref{fig:compatibility}, the two processes interact only when DRRS signal injection happens after the checkpoint barrier is forwarded by the predecessor but not yet processed by scaling instances.
In this case, to avoid interference, the checkpoint barrier is treated as an additional constraint on DRRS signal propagation.
When the checkpoint barrier is in the output buffer (Fig~\ref{compatibility:a}), redirection concludes at the barrier, which then functions as an integrated signal combining three barriers.
Upon receiving  this signal, the instance responds sequentially to the checkpoint barrier, trigger barrier, and confirm barrier.
Conversely, if the checkpoint barrier is in the input buffer (Fig~\ref{compatibility:b}), the confirm barrier operates as usual, while the trigger barrier is integrated into the checkpoint barrier.}
Moreover, to handle potential scaling failures, DRRS incorporates scaling-related states, such as subscale progress and in-transit data, within snapshots.

\section{Evaluation}
\label{evaluation}
In this section, we present an empirical evaluation of our DRRS prototype through a systematic three-stage methodology,
which progressively validates its effectiveness from core mechanism verification in controlled environments to sensitivity analysis under production-like conditions.
We focuses on the following research questions:

\begin{enumerate}
  \item \textbf{Fundamental Effectiveness}: How does DRRS outperform SOTA on-the-fly scaling approaches (Megaphone\cite{hoffmann2019megaphone} and Meces\cite{gu2022meces})? (Section \ref{sec:controlled_environment})
  \item \textbf{Design Rationale Validation: 
  What are the quantitative contributions of DRRS's three core mechanisms to overall system performance?} (Section \ref{sec:ablation_study}) 
  \item \textbf{Practical Sensitivity}: How does DRRS perform on a multi-machine cluster, with varying state sizes, input rates, and workload skewness? (Section \ref{sec:sensitivity_analysis})
\end{enumerate}

\subsection{Experimental Setup}
\label{sec:experimental_setup}
\noindent \textbf{Hardware and Software}:
The DRRS prototype is implemented on Apache Flink 1.17.0 \cite{flink_release_1_17}, 
and the evaluation uses two experimental setups, including a single-machine Dockerized environment and a multi-machine Docker Swarm cluster.
The single-machine setup runs on a server equipped with dual Intel Xeon Gold 5218 processors (2.30GHz) and 256GB of memory, isolating the system from non-system-design factors such as network jitter or hardware heterogeneity.
The Swarm cluster consists of three additional servers:
two with dual Intel Xeon Silver 4210 processors (2.20GHz) and 128GB of memory, 
and one with dual Intel Xeon Gold 6230 processors (2.10GHz) and 256GB of memory.
This 4-node heterogeneous cluster is designed to assess DRRS's performance in a relatively production-like environment.
All servers run Ubuntu 18.04.6 LTS, OpenJDK 11, and Docker Engine v24.0.2 as the core software components, and are connected via Gigabit Ethernet (1 Gbps).

\noindent \textbf{Workloads and Metrics}: 
We use three distinct workloads to comprehensively assess DRRS.

First, we utilize NEXMark\cite{tucker2008nexmark}, a synthetic benchmark that simulates a high-throughput auction system.
We specifically focus on Queries 7 and 8 (Q7 and Q8) due to their wide use in related research\cite{gu2022meces,song2023sponge,del2020rhino,hoffmann2019megaphone,del2022rethinking}.
To ensure consistent scaling behavior in our experiments, we use Sliding Window operators instead of Tumbling Window operators, as the latter can introduce significant instability in scaling performance due to their periodic state accumulation and batch processing nature.

Second, we use a real-world workload derived from the Twitch streaming platform \cite{rappaz2021recommendation}, 
which is a seven-operator pipeline that analyzes viewer engagement patterns to compute loyalty scores.
To ensure experimental feasibility while preserving the real-world data characteristics, we randomly select a one-fifth subset of the dataset, comprising approximately 4 million events, and compress the event timestamps into a 1,000-second window.

Additionally, for performance evaluation under varying operational conditions, we develop a 3-operator job---a generator, a keyed aggregator, and a sink---as a straightforward custom workload, given that the major overhead of on-the-fly scaling occurs only in the scaling operator and its predecessors. 
This configurable stateful workload allows for adjustable parameters, including state size, input rate, and workload skewness.

We employ the NEXMark and Twitch workloads, both sourcing system inputs through Apache Kafka 3.8.0 running on a separate server, to evaluate the fundamental effectiveness and design rationale of DRRS.
The custom workload, utilizing internal data generators to better capture scaling-induced latency variations, serves our sensitivity analysis under cluster deployment scenarios.

\begin{figure*}[t]
  \centering
  \begin{subfigure}[t]{\linewidth}
    \centering
    \includegraphics[width=\linewidth]{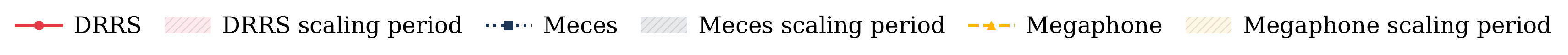}
  \end{subfigure}

  \centering
  \begin{subfigure}[t]{0.32\linewidth}
    \centering
      \includegraphics[height=4.2cm]{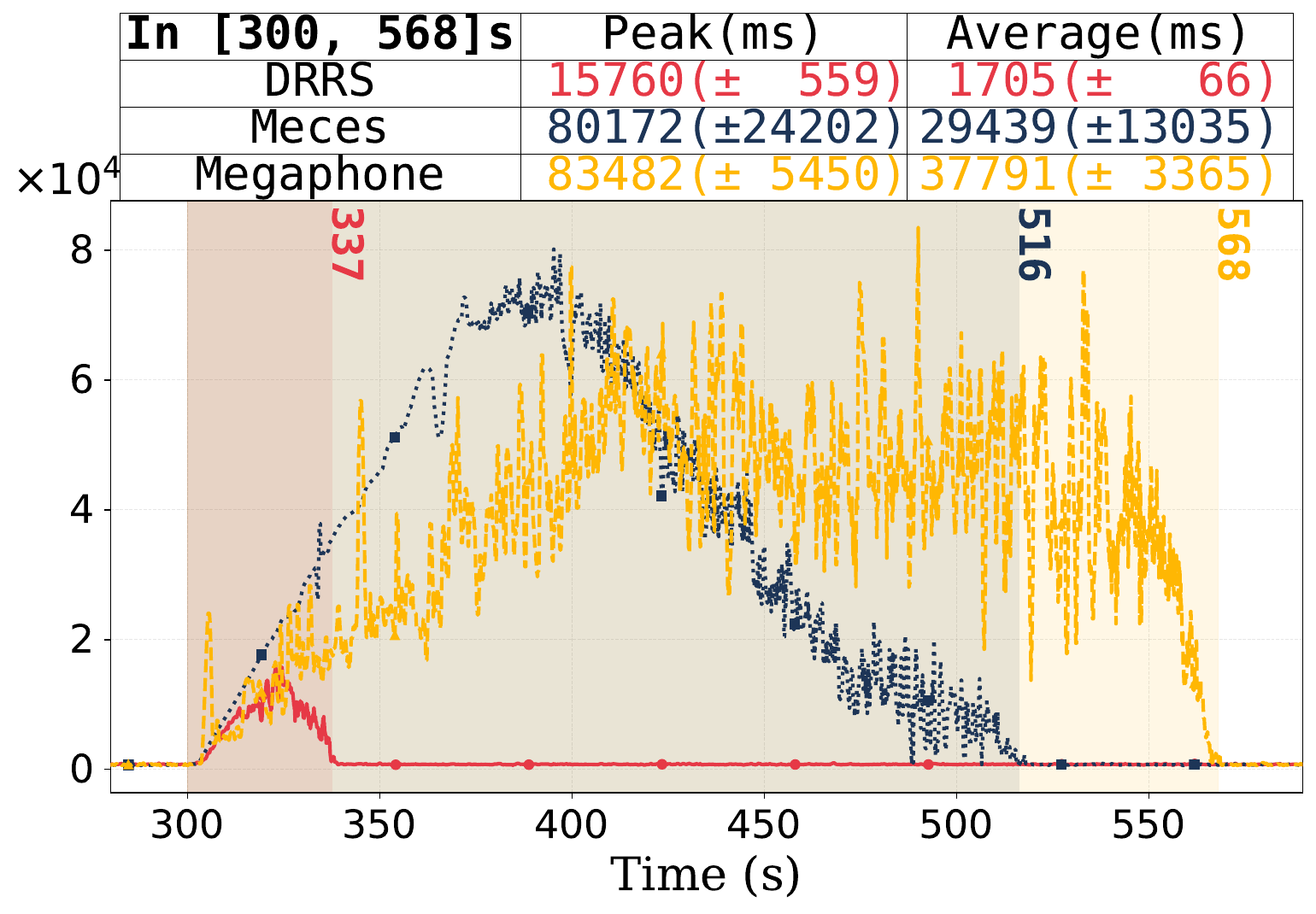}
    \caption{Q7}
    \label{latency:q7}
  \end{subfigure}
    \hfill
  \begin{subfigure}[t]{0.32\linewidth}
    \centering
    \includegraphics[height=4.2cm]{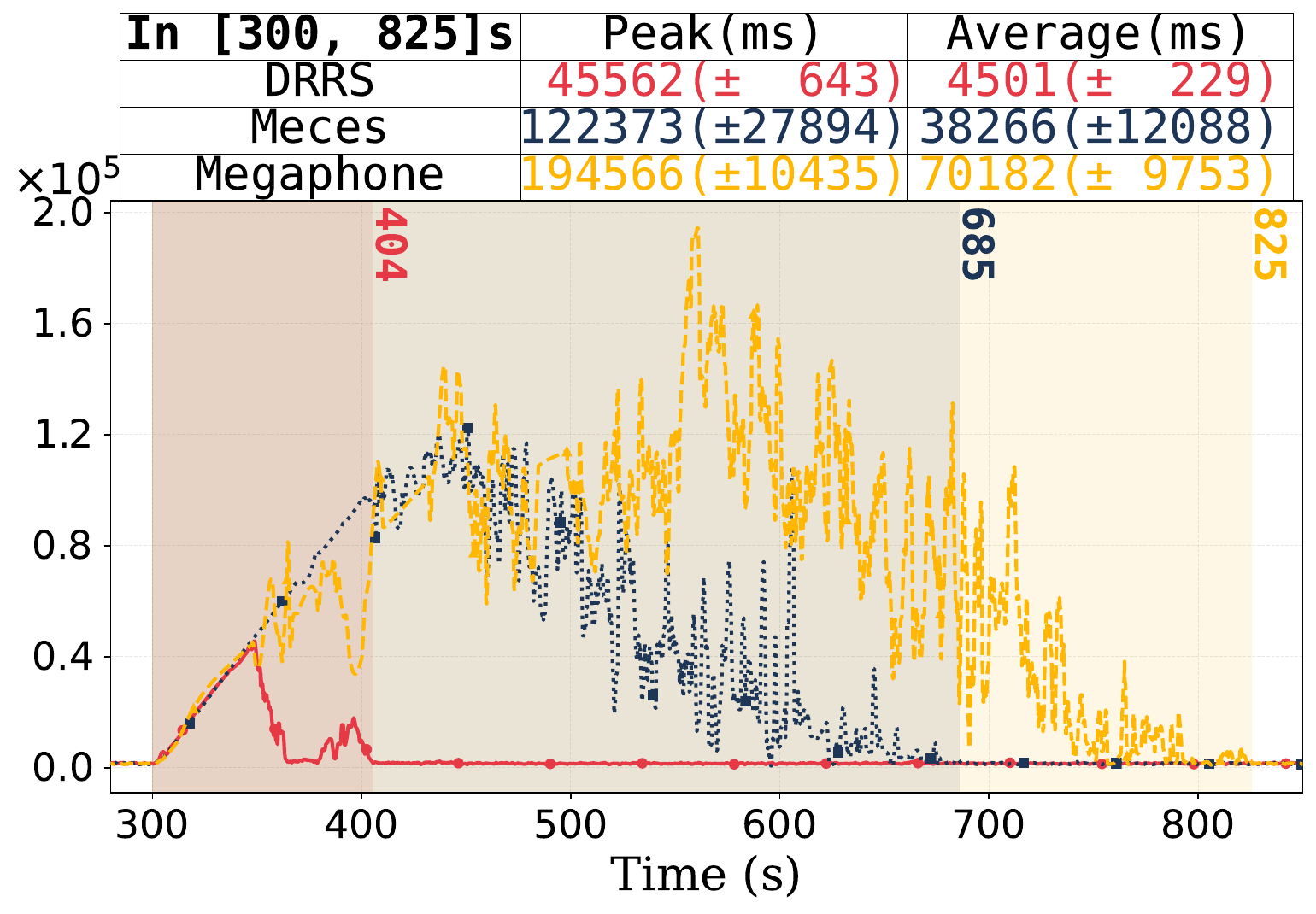}
    \caption{Q8}
    \label{latency:q8}
  \end{subfigure}
    \hfill
  \begin{subfigure}[t]{0.32\linewidth}
    \centering
    \includegraphics[height=4.2cm]{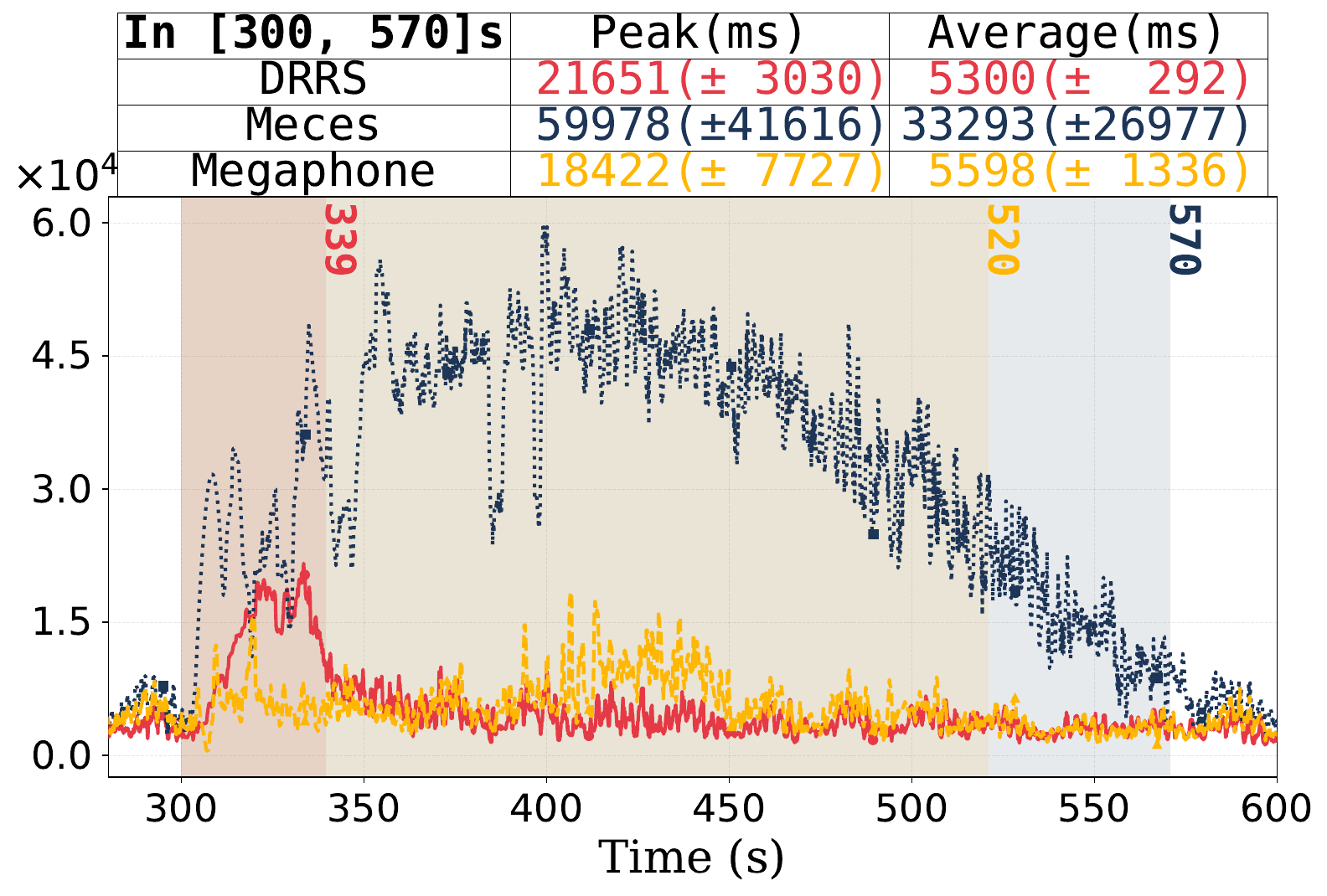}
    \caption{Twitch}
    \label{latency:twitch}
  \end{subfigure}
  
  \caption{{Comparison in terms of End-to-End Latency (ms).}}
  \label{fig:latency}
\end{figure*}

Our evaluation focuses on 
\textbf{end-to-end latency} and \textbf{throughput} as key performance metrics.
We measure the end-to-end latency using periodically inserted latency markers that flow through the system as regular records, while bypassing windowing operators to avoid windowing effects on measurements.
This latency measurement includes the Kafka transit time and the additional latency introduced by backpressure's impact on the Source operators.
Throughput is measured by the output rate of the source operators, accounting for both Kafka consumption and internal data generation, providing an accurate assessment of the system's processing capacity.

\noindent \textbf{Baseline Implementations}:
Our implementations of Megaphone and Meces within Apache Flink 1.17.0 maintain fidelity to their original architectures while ensuring consistent experimental conditions for fair comparison.
While Megaphone was originally designed for specific SPEs~\cite{murray2013naiad} and required additional operators for fluid migration,
our implementation achieves comparable short propagation paths through predecessor injection, which aligns with the non-generalized separation of control and data planes in the original design.
We also preserve Megaphone's characteristic time-stamp-driven scaling process by employing the Naive Division strategies detailed in Section~\ref{sec:subscale_division}.
The original Meces implementation~\cite{meces_artifact} depends on external Redis clusters for state transmission.
We integrate its core features---Fetch-on-Demand and Hierarchical State Organization---within Apache Flink's framework.
This integration ensures that all Meces's scaling operations comply with Flink's native mechanisms and operate under the same system conditions as DRRS and Megaphone, 
particularly regarding resource management and Netty-based transmission protocols.

The DRRS implementation adds a 200-item buffer (\textasciitilde200KB per scaling instance) for pre-serializing records to support Record Scheduling, 
which is also added to Megaphone. However, Meces with this buffer retrieves states more aggressively, 
leading to more frequent back-and-forth migrations and a subsequent performance decline. 
Thus, we exclude this buffer in Meces.
Additionally, since fluid migration is used across all the three systems, we maintain a consistent configuration with key-group serving as the atomic migration unit.

\subsection{Fundamental Effectiveness}
\label{sec:controlled_environment}

\begin{figure}[tb]
    \centering

        \begin{subfigure}[t]{0.32\linewidth}
            \centering
            \includegraphics[height=2.9cm, trim=5 0 5 0, clip]{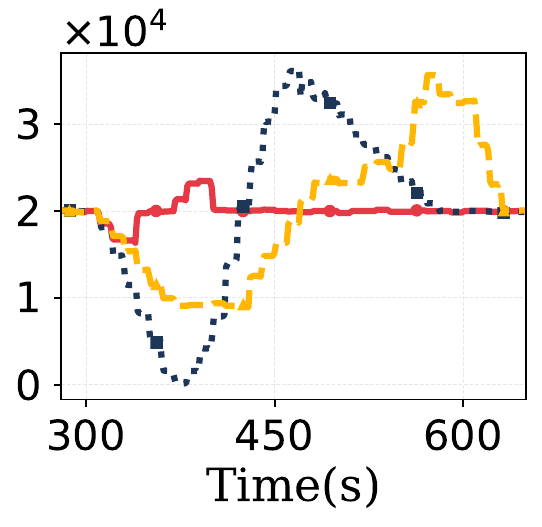}
            \caption{Q7}
            \label{throughput:q7}
        \end{subfigure}%
        \hfill
        \begin{subfigure}[t]{0.345\linewidth}
            \centering
            \includegraphics[height=2.9cm, trim=5 0 5 0, clip]{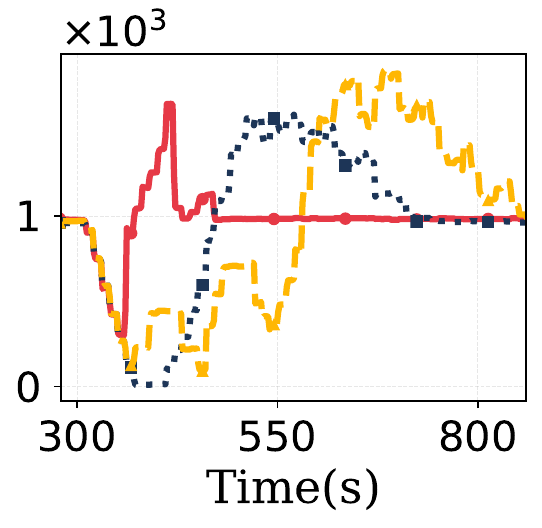}
            \caption{Q8}
            \label{throughput:q8}
        \end{subfigure}%
        \hfill
        \begin{subfigure}[t]{0.32\linewidth}
            \centering
            \includegraphics[height=2.9cm, trim=5 0 5 0, clip]{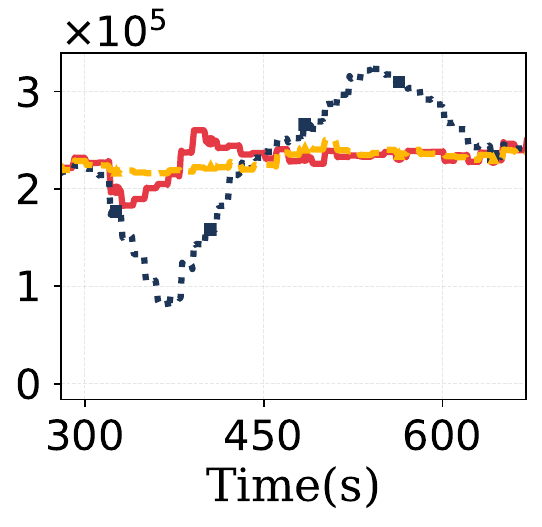}
            \caption{Twitch}
            \label{throughput:twitch}
        \end{subfigure}
    \caption{Comparison in terms of Throughput (records/s).}
    \label{fig:throughput}
\end{figure}

We first evaluate the fundamental effectiveness of DRRS by comparing it to Megaphone and Meces under identical deployment settings (1 CPU core/8GB memory per container).
Each experiment progresses through three phases: 
1) a 300-second warm-up for steady-state establishment;
2) a scaling operation expanding bottleneck operators from 8 to 12 instances, migrating 111 key-groups out of 128 with uniform re-partitioning;
3) a post-scaling stabilization period.
To validate system behavior across diverse scenarios, we use three workloads: NEXMark Q7, Q8, and Twitch, each with distinct characteristics.

Q7 features high input rates (20K tps) with a 10-second window size and 500ms sliding interval, maintaining state sizes approaching 800MB,
while Q8 operates at lower input rates (1K tps) with a 40-second window size and 5-second sliding interval, managing larger state sizes (\textasciitilde3GB).
The Twitch workload complements these synthetic benchmarks by preserving authentic data patterns,
facilitating natural state accumulation through continuous processing and reaching total state sizes of approximately 500MB when scaling begins.
All measurements are derived from multiple experimental iterations to maintain statistical validity and ensure reproducibility of results.

Fig.~\ref{fig:latency} and~\ref{fig:throughput} demonstrate that the system performance during scaling exhibits similar patterns across all tests:
Upon scaling begins, latency rapidly rises to a peak before gradually decreasing, while throughput initially drops and then increases to a higher level, eventually stabilizing.
This overcompensation in throughput results from the backpressure generated during scaling,
which causes records to buffer and subsequently flush upon scaling completion, thereby compensating throughput to a higher level.
This overload subsequently triggers new upstream backlogs, initiating a cycle of system oscillations that continues until the system reaches a steady state.
This process contributes to significant latency fluctuations or even rebounds during its descent phase.
Therefore, we define the scaling period as the interval from the initial scaling operation until latency re-stabilizes---specifically, when latency keeps within 110\% of the pre-scaling level for 100 seconds. 
We use the longest observed scaling period among all three methods as the statistical basis for subsequent analysis.

\begin{figure}[tb]
  \centering
  \begin{subfigure}[t]{\linewidth}
    \centering
    \includegraphics[width=0.8\linewidth, trim=40 50 60 50, clip]{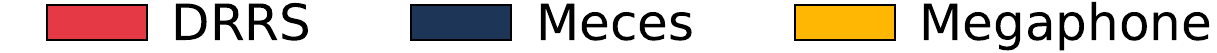}
  \end{subfigure}

    \centering
    \begin{subfigure}[t]{0.497\linewidth}
        \centering
        \includegraphics[width=\linewidth]{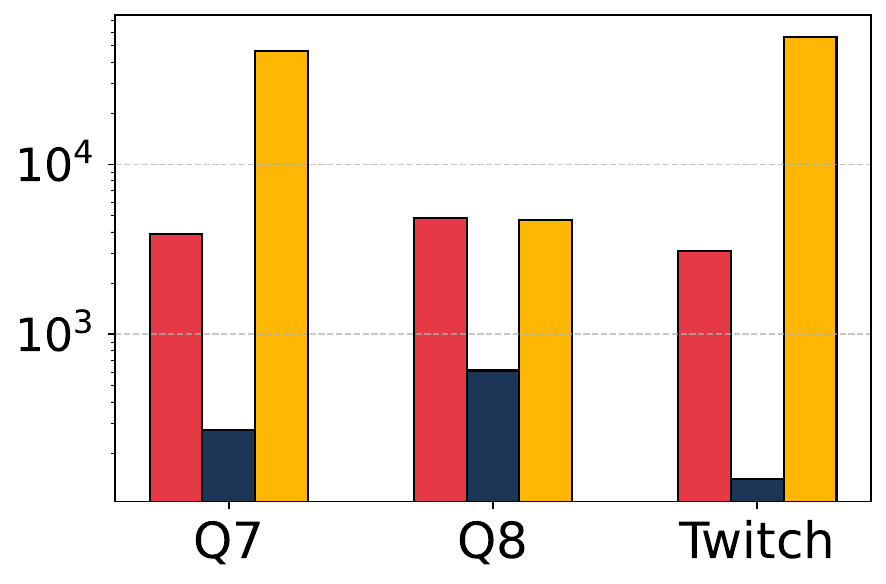}
        \caption{Prop. Delay (ms)}
        \label{fig:propagation}
    \end{subfigure}%
    \hfill
    \begin{subfigure}[t]{0.5\linewidth}
        \centering
        \includegraphics[width=\linewidth]{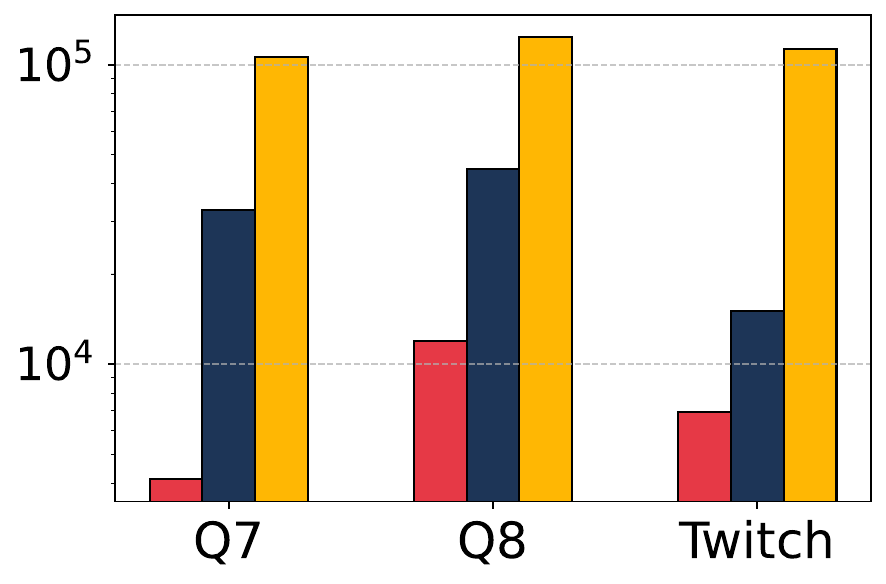}
        \caption{Dependency Overhead (ms)}
        \label{fig:depends}
    \end{subfigure}%
    \caption{Comparison in terms of Cumulative Propagation Delay (sum of time intervals between signal injection and first state migration) and Average Dependency-Related Overhead (average time intervals from signal injection to state migration across states).}
    \label{fig:combind}
\end{figure}

\begin{figure}[tb]
    \centering
\begin{subfigure}[t]{0.32\linewidth}
        \centering
        \includegraphics[height=2.9cm, trim=5 0 5 0, clip]{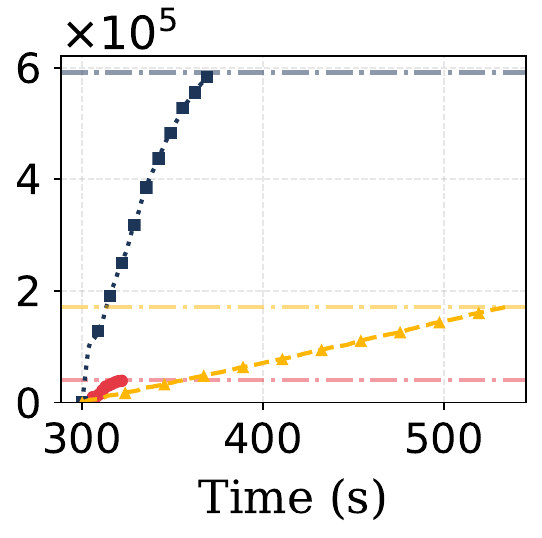}
        \caption{Q7}
        \label{suspend:q7}
    \end{subfigure}%
    \hfill
    \begin{subfigure}[t]{0.345\linewidth}
        \centering
        \includegraphics[height=2.9cm, trim=5 0 5 0, clip]{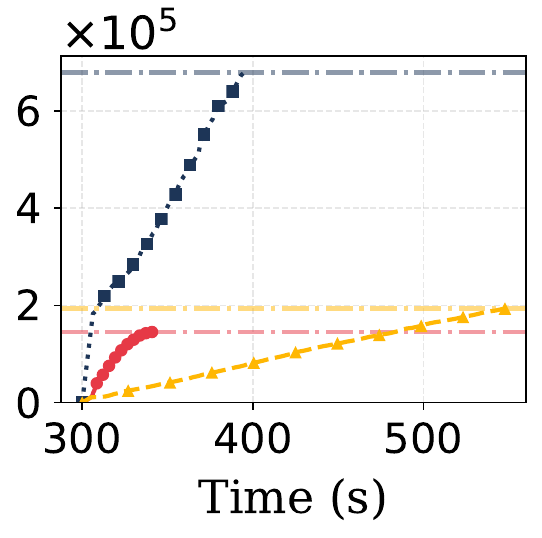}
        \caption{Q8}
        \label{suspend:q8}
    \end{subfigure}%
    \hfill
    \begin{subfigure}[t]{0.32\linewidth}
        \centering
        \includegraphics[height=2.9cm, trim=5 0 5 0, clip]{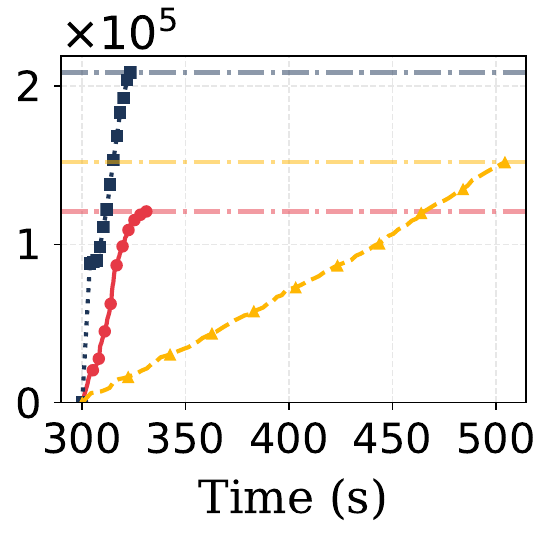}
        \caption{Twitch}
        \label{suspend:twitch}
    \end{subfigure}
    \caption{Comparison in Cumulative Suspension Time (ms).}
    \label{fig:suspend}
\end{figure}

The experiments demonstrate DRRS's substantial improvements in both latency reduction and system stability, as shown in Fig.~\ref{latency:q7} and \ref{latency:q8}.
In NEXMark Q7, compared to Megaphone and Meces, DRRS reduces peak latency by 81.1\% and 80.3\%, average latency by 95.5\% and 94.2\%, and scaling time by 86\% and 82.7\%.
This advantage persists in NEXMark Q8, with the reductions of 76.6\%/62.8\% (peak), 93.6\%/88.2\% (average), and 80.1\%/72.8\% (time), respectively.
In Twitch (Fig.~\ref{latency:twitch}), while Megaphone's conservative migration yields comparable peak and average latencies,
DRRS still achieves an 82.2\% reduction in scaling time compared to Megaphone, and consistently outperforms Meces across all the metrics.
The throughput data in Fig.~\ref{fig:throughput} further highlights DRRS's ability to ensure limited fluctuations and a rapid return to stable levels across all workloads.
To systematically analyze these performance gains,
we discuss the limitations of baseline methods through three key metrics: 
cumulative propagation latency, average dependency overhead, and cumulative suspension time (Fig.~\ref{fig:combind} and \ref{fig:suspend}).

The timestamp-driven nature of Megaphone enables a slow growth in suspension time across all workloads (Fig.~\ref{fig:suspend}).
However, Megaphone necessitates extensive synchronization operations
and enforces mandatory input blocking for alignment,
creating a strict linear dependency relationship between migration units.
Consequently, as evidenced in Fig. \ref{fig:combind},
Megaphone's cumulative propagation delay and average dependency overhead significantly surpass those of other methods.
The impact leads to extended scaling durations, reaching up to 7.24 times longer than DRRS in the Q7 workload. 
It diminishes the system's ability to respond to load fluctuations.
The sequential dependency further compounds these issues by introducing cascading performance degradation during extended scaling operations.

In comparison to Megaphone, DRRS with the Decoupling and Re-routing mechanism decouples synchronization signals and eliminates alignments, 
ensuring manageable propagation overhead and allowing migration to proceed smoothly even in the presence of performance degradation. 
The Subscale Division mechanism reduces dependency overhead, 
significantly shortening scaling duration and facilitating the rapid activation of new nodes.
This prevents the accumulation of prolonged performance degradation and the cascading amplification effects.

While Meces achieves the lowest cumulative propagation overhead through its single-synchronization design (Fig.~\ref{fig:combind}), its Fetch-on-Demand mechanism introduces a critical back-and-forth migration issue, significantly increasing suspension time as shown in Fig.~\ref{fig:suspend}. 
Meces's support for early migration requires both migration in/out instances to access records simultaneously, leading to substantial fetch conflicts in spite of its Hierarchical State Organization. 
In the Q7 workload, this fetch involves a total of 55 sub-key-groups, with one sub-key-group being migrated 6.25 times on average, and up to 46 times. 
Moreover, these fetch conflicts lead to significant instabilities, as the Fetch-on-Demand requirement is closely tied to the system's real-time input patterns.

\begin{figure}[tb]
    \centering
        \includegraphics[width=\linewidth]{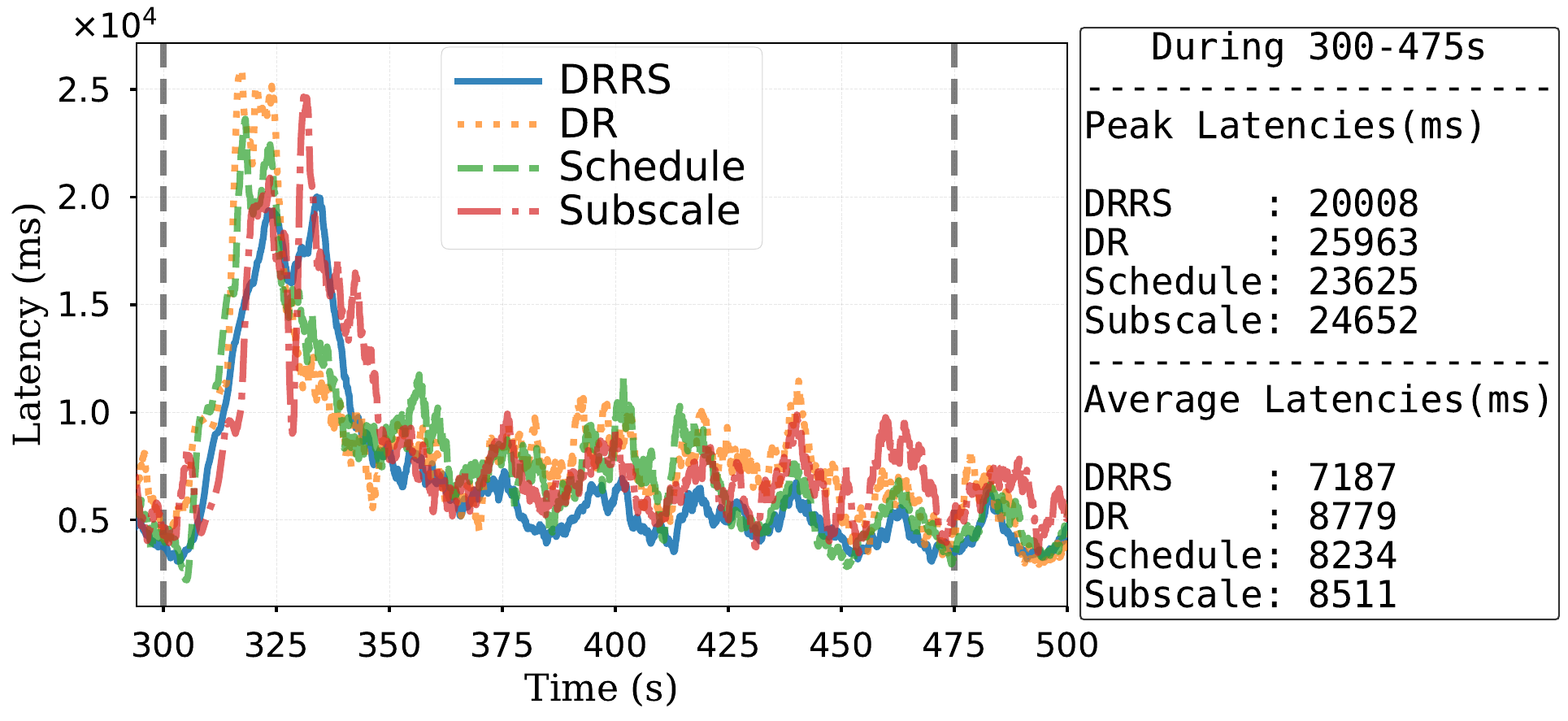}
    \caption{Impact of DRRS Mechanisms on Latency.}
    \label{fig:ablation}
\end{figure}

In comparison to Meces, DRRS supports even earlier migration by prioritizing trigger barriers and uses a stream-operation approach to address the simultaneous requirement problem, which is more lightweight than Meces's state-operation approach.
With Rerouting, DRRS not only ensures consistency of execution order beyond the exactly-once guarantee provided by Meces, but also achieves lower transmission overhead.
Moreover, Decoupling and Rerouting limits the number of records that need to be re-routed.
Meanwhile, DRRS reduces suspension through Record Scheduling as shown in Fig.~\ref{fig:suspend}.

\subsection{Design Rationale Validation}
\label{sec:ablation_study}

Having established DRRS's fundamental advantages, we now conduct an isolation test using the Twitch workload and the same experimental setup to quantify the individual contributions of DRRS's core designs. 
Fig.~\ref{fig:ablation} presents comparisons across four system variants: the complete DRRS system and three variants, each only including one core design: Decoupling and Re-routing (\textit{DR}), Record Scheduling (\textit{Schedule}), and Subscale Division (\textit{Subscale}).

Specifically, the integrated DRRS system achieves the lowest peak and average latencies. In contrast, \textit{DR} exhibits the most pronounced performance degradation, with peak latency increasing by 30\% and average latency by 22\%, indicating that propagation delay is not the primary bottleneck in the Twitch workload. 
When considered in isolation, Record Scheduling and Subscale Division increase peak/average latency by 18\%/15\% and 23\%/18\%, respectively. 
Furthermore, \textit{Subscale} undergoes the largest fluctuations, largely due to multiple synchronizations interfering with one another.
These results demonstrate that while each mechanism individually contributes to the overall optimization, they also complement one another to produce synergistic benefits.

\subsection{Sensitivity Analysis}
\label{sec:sensitivity_analysis}

To thoroughly evaluate the performance of DRRS under more realistic deployment conditions, we conduct a sensitivity analysis using the Swarm cluster. 
The experimental setup incorporates three major modifications:
(1) expanding the number of key-groups to 256,
(2) scaling operator instances from 25 to 30 (triggering migration of 229 key-groups),
and (3) increasing container memory allocation to 20GB.  
We evaluate the impacts of 
input rates (5K-20K tps), total state sizes (5-30GB), and workload skewness modeled through Zipf distributions (skewness parameters in [0.0, 0.5, 1.0, 1.5]).
Throughput deviation from the input rates serves as the performance metric (collected over a 10-minute period), 
chosen over latency measurements because the queuing backlogs under high skewness could compromise latency measurement reliability.

\begin{figure}[tb]
    \centering
    \includegraphics[width=\linewidth]{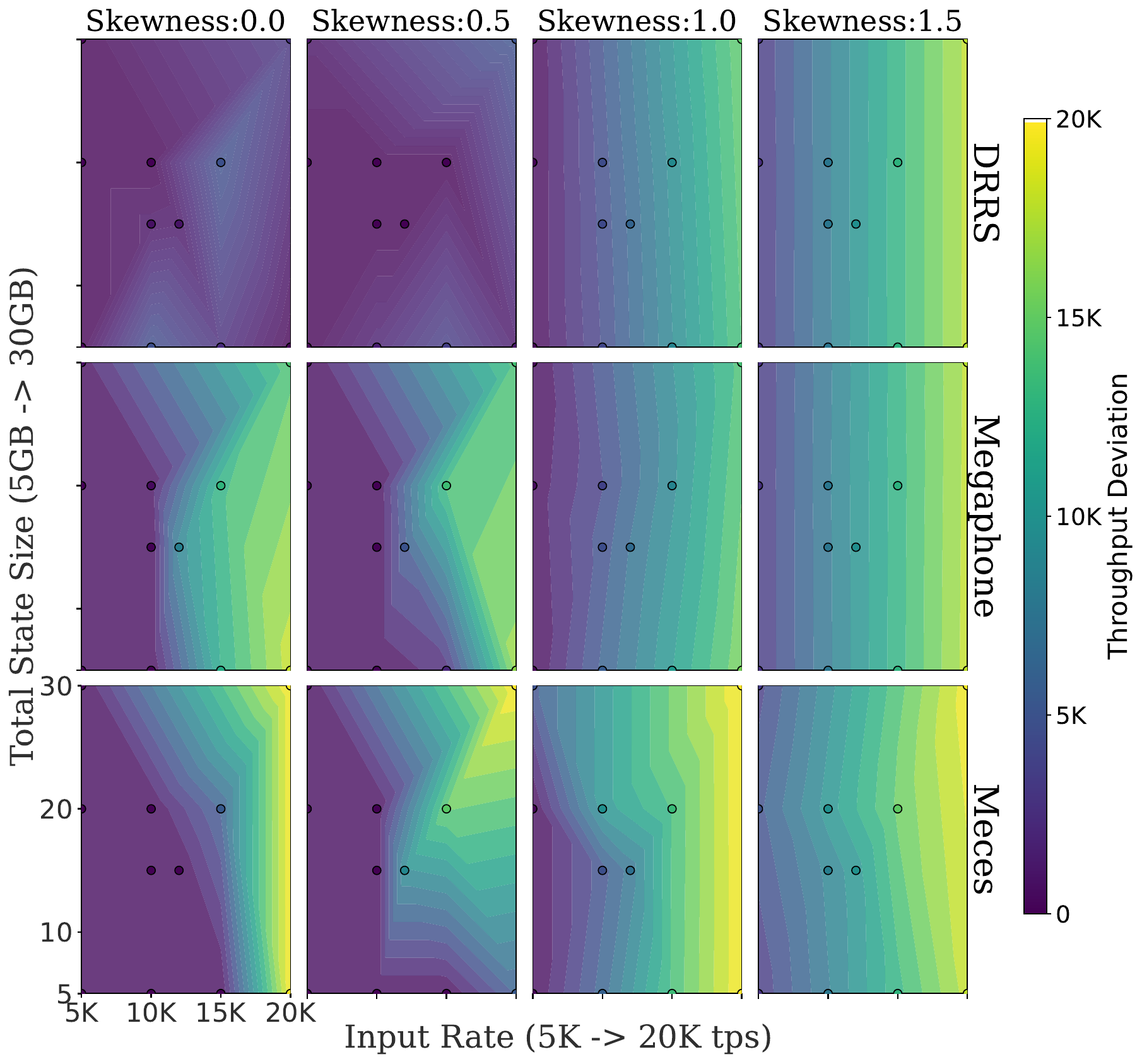}
    \caption{Sensitivity Analysis of Throughput Deviation.}
    \label{fig:sensitivity}
\end{figure}

Fig.~\ref{fig:sensitivity} presents the results in a three-dimensional visualization: the x-axis for input rates, y-axis for state sizes, and the panel arrangement corresponds to skewness levels. 
Here we use <input rate, state size, skewness> to identify a specific scenario.
The color gradient illustrates throughput deviation, with purple for lower deviation (better performance) and yellow for higher deviation (worse performance).
The results largely align with theoretical expectations, showing progressive performance degradation as state sizes, input rates, or skewness levels increase. 
However, some anomalies are observed in Megaphone and Meces, where systems with higher skewness (1.5) can outperform those with lower skewness (0.0). This phenomenon is closely related to the scaling process.

For Meces, performance is significantly influenced by the frequency of fetch-on-demand operations affected by real-time system inputs and state migration progress, leading to considerable instability. We observe that at the configuration <20k, 5GB, 1.0>, Meces completes scaling faster than at <20k, 5GB, 0.0>, thereby returning to a stable state more quickly, which results in lower throughput deviation.
In contrast, due to Megaphone's prolonged scaling time, it is unable to complete the entire migration process within the measurement time, even in the 0.0 skewness scenario. 
In such cases, because of the single-instance involvement, the other instances continue to operate, maintaining relatively high throughput.
This situation worsens at higher skewness levels, resulting in even less state being migrated, though not reflected by the system throughput.
Once skewness reaches a higher level, processing pauses caused by migration in a single instance can trigger significant backpressure in all upstream instances, ultimately returning the system to the expected growth behavior.

Under the amplified performance volatility inherent in distributed deployments, DRRS demonstrates consistent superiority across all parameter combinations.
The advantage is particularly pronounced at greater state sizes and input rates, 
where DRRS achieves up to 89\% higher throughput than baseline methods in 30GB with 20K tps,
showcasing its robustness and adaptability to varying operational conditions.

\section{Related Work}
\label{relatedwork}
Scalability is a key feature of modern SPEs, enabling them to effectively handle system dynamics such as workload fluctuations and resource availability variations. This capability has been a focal point of research for many years \cite{cherniack2003scalable,shah2003flux, cardellini2016elastic, lin2015scalable,gedik2013elastic,volnes2023migrate}.
While on-the-fly scaling approaches, as aforementioned, must account for both state and stream management during scaling, other related research can be categorized into two primary domains: state-centric and stream-centric studies.

State-centric studies focus on state management mechanisms tailored to scaling requirements, with the aim of reducing overhead during state migration. 
Checkpoint-assisted scaling\cite{castro2013integrating,carbone2017state,noghabi2017samza} has garnered significant attention due to its wide applicability and the potential to share overhead with fault tolerance mechanisms.
In addition to techniques like unaligned\cite{apache_flink_checkpointing} and incremental\cite{richter_ward_incremental_checkpointing} checkpointing that directly reduce checkpointing overhead, several approaches\cite{madsen2015dynamic,mao2021trisk,wu2015chronostream} employ task-level checkpoint management and prioritize the storage of checkpoints on the nodes where new instances would be deployed, thereby enhancing data locality and reducing network I/O during state migration.
In addition, proactive migration \cite{del2020rhino} performs regular state migration in non-scaling periods to reduce overhead during scaling. However, it can degrade performance and require additional resources during regular operation. 
Other methods have been proposed for particular applications, such as state shedding \cite{volnes2022travel} that selectively eliminates non-critical states during scaling to reduce migration overhead, which is only suitable if applications can tolerate approximate results.

Stream-centric studies mainly focuses on improving stream operations during scaling to reduce overhead.
Parallel-track approaches \cite{rajadurai2018gloss,gulisano2012streamcloud,ottenwalder2013migcep} achieve zero suspension by duplicating input streams and migrating state replicas. 
Specifically for window-based operators, window recreation\cite{pham2017uninterruptible,madsen2016enorm,luthra2018tcep} further enables implicit state migration without any network I/O overhead. However, these methods introduce additional computational overhead due to record duplication and reprocessing.

Several studies on scaling decisions incorporate scaling overhead into scheduling as an optimization objective. For example, key-partition distributions are calculated using refined algorithms like consistent hashing to reduce the number of migration units\cite{heinze2014latency,fang2018distributed,gulisano2012streamcloud}; new instance placement is optimized based on state migration costs\cite{wang2019elasticutor,madsen2015dynamic}.
Some other approaches consider scaling optimization from a resource perspective. The adoption of lightweight computing resources\cite{spenger2022portals,jain2020splitserve}, such as stateful serverless functions\cite{sreekanti2020cloudburst}, enables a more efficient response to rapid load fluctuations\cite{song2023sponge}.

In comparison, DRRS as an on-the-fly scaling method manages both states and streams without extra resource, and is orthogonal to scaling decisions.
Meanwhile, DRRS can work with existing solutions to address a broader range of scaling requirements. For instance, DRRS can leverage the record scheduling mechanism of Rhino \cite{del2020rhino} in conjunction with periodic state snapshots, enhancing the applicability in extensive state management.

\section{Conclusion}
\label{conclusion}
This paper presents DRRS, a flexible and efficient on-the-fly scaling method to minimize performance degradation during scaling in stateful SPEs.
DRRS incorporates three mechanisms: Decoupling and Re-routing to reduce synchronization overhead, Reordering to mitigate processing suspensions, and Subscale Division to decrease dependency-related migration overhead.
We implement DRRS within Apache Flink as a plugin, enabling seamless integration with existing applications and no disruption during non-scaling periods.
Our evaluation across diverse workloads demonstrates that, compared to state-of-the-art mechanisms, DRRS significantly improves system performance during scaling.

In future work, we plan to integrate DRRS with advanced scaling decision-making algorithms to fully leverage its fine-grained control and flexibility, enhancing the adaptivity of stateful stream processing systems.



\bibliographystyle{IEEEtran}
\bibliography{arxiv}

\end{document}